\documentclass[twoside,twocolumn,9pt]{article}
\pdfoutput=1
\usepackage{extsizes}
\usepackage[super,sort&compress,comma]{natbib} 
\usepackage[version=3]{mhchem}
\usepackage[left=1.5cm, right=1.5cm, top=1.785cm, bottom=2.0cm]{geometry}
\usepackage{balance}
\usepackage{times,mathptmx}
\usepackage{sectsty}
\usepackage{notes2bib}
\usepackage{graphicx} 
\usepackage{lastpage}
\usepackage[format=plain,justification=justified,singlelinecheck=false,font={stretch=1.125,small,sf},labelfont=bf,labelsep=space]{caption}
\usepackage{float}
\usepackage{fancyhdr}
\usepackage{fnpos}
\usepackage[english]{babel}
\addto{\captionsenglish}{%
  \renewcommand{\refname}{Notes and references}
}
\usepackage{array}
\usepackage{droidsans}
\usepackage{charter}
\usepackage[T1]{fontenc}
\usepackage[usenames,dvipsnames]{xcolor}
\usepackage{setspace}
\usepackage[compact]{titlesec}
\usepackage{hyperref}

\usepackage{epstopdf}

\definecolor{cream}{RGB}{222,217,201}

\begin{document}

\pagestyle{fancy}
\thispagestyle{plain}
\fancypagestyle{plain}{

\fancyhead[C]{\includegraphics[width=18.5cm]{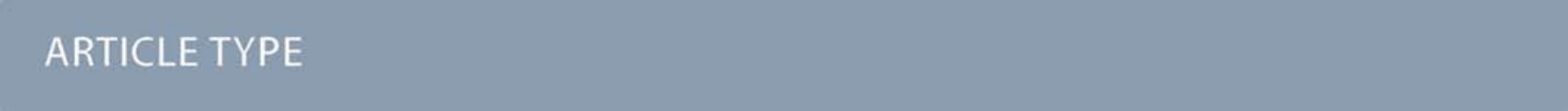}}
\fancyhead[L]{\hspace{0cm}\vspace{1.5cm}\includegraphics[height=30pt]{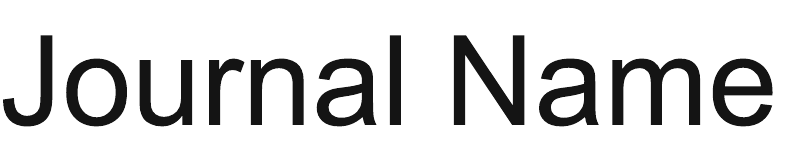}}
\fancyhead[R]{\hspace{0cm}\vspace{1.7cm}\includegraphics[height=55pt]{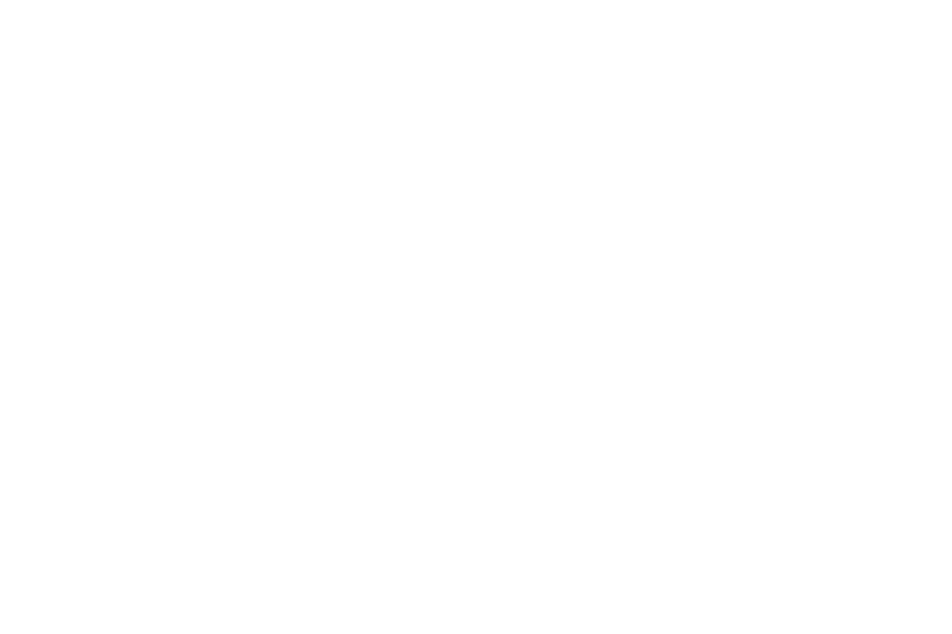}}
\renewcommand{\headrulewidth}{0pt}
}

\makeFNbottom
\makeatletter
\renewcommand\LARGE{\@setfontsize\LARGE{15pt}{17}}
\renewcommand\Large{\@setfontsize\Large{12pt}{14}}
\renewcommand\large{\@setfontsize\large{10pt}{12}}
\renewcommand\footnotesize{\@setfontsize\footnotesize{7pt}{10}}
\makeatother

\renewcommand{\thefootnote}{\fnsymbol{footnote}}
\renewcommand\footnoterule{\vspace*{1pt}%
\color{cream}\hrule width 3.5in height 0.4pt \color{black}\vspace*{5pt}} 
\setcounter{secnumdepth}{5}

\makeatletter 
\renewcommand\@biblabel[1]{#1}            
\renewcommand\@makefntext[1]%
{\noindent\makebox[0pt][r]{\@thefnmark\,}#1}
\makeatother 
\renewcommand{\figurename}{\small{Fig.}~}
\sectionfont{\sffamily\Large}
\subsectionfont{\normalsize}
\subsubsectionfont{\bf}
\setstretch{1.125} 
\setlength{\skip\footins}{0.8cm}
\setlength{\footnotesep}{0.25cm}
\setlength{\jot}{10pt}
\titlespacing*{\section}{0pt}{4pt}{4pt}
\titlespacing*{\subsection}{0pt}{15pt}{1pt}

\fancyfoot{}
\fancyfoot[LO,RE]{\vspace{-7.1pt}\includegraphics[height=9pt]{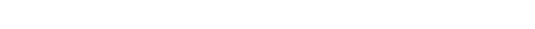}}
\fancyfoot[CO]{\vspace{-7.1pt}\hspace{13.2cm}\includegraphics{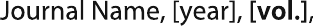}}
\fancyfoot[CE]{\vspace{-7.2pt}\hspace{-14.2cm}\includegraphics{head_foot/RF}}
\fancyfoot[RO]{\footnotesize{\sffamily{1--\pageref{LastPage} ~\textbar  \hspace{2pt}\thepage}}}
\fancyfoot[LE]{\footnotesize{\sffamily{\thepage~\textbar\hspace{3.45cm} 1--\pageref{LastPage}}}}
\fancyhead{}
\renewcommand{\headrulewidth}{0pt} 
\renewcommand{\footrulewidth}{0pt}
\setlength{\arrayrulewidth}{1pt}
\setlength{\columnsep}{6.5mm}
\setlength\bibsep{1pt}

\makeatletter 
\newlength{\figrulesep} 
\setlength{\figrulesep}{0.5\textfloatsep} 

\newcommand{\topfigrule}{\vspace*{-1pt}%
\noindent{\color{cream}\rule[-\figrulesep]{\columnwidth}{1.5pt}} }

\newcommand{\botfigrule}{\vspace*{-2pt}%
\noindent{\color{cream}\rule[\figrulesep]{\columnwidth}{1.5pt}} }

\newcommand{\dblfigrule}{\vspace*{-1pt}%
\noindent{\color{cream}\rule[-\figrulesep]{\textwidth}{1.5pt}} }

\makeatother

\twocolumn[
  \begin{@twocolumnfalse}
\vspace{3cm}
\sffamily
\begin{tabular}{m{4.5cm} p{13.5cm} }

\includegraphics{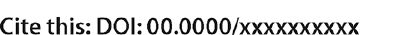} & \noindent\LARGE{\textbf{Phase separation of mixtures after a second quench: composition heterogeneities}} \\
\vspace{0.3cm} & \vspace{0.3cm} \\

 & \noindent\large{Pablo de Castro$^{\ast a, b}$ and Peter Sollich$^{a, c}$} \\

\includegraphics{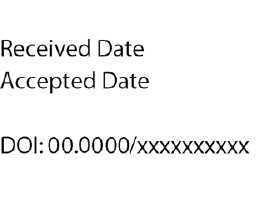} & \noindent\normalsize{We investigate binary mixtures undergoing phase separation after a \textit{second} (deeper) temperature quench into two- and three-phase coexistence regions. The analysis is based on a lattice theory previously developed for gas-liquid separation in generic mixtures. Our previous results, which considered an arbitrary number of species and a single quench, showed that, due to slow changes in composition, dense colloidal mixtures can phase-separate in two stages. Moreover, the denser phase contains long-lived composition heterogeneities that originate as the interfaces of shrunk domains. Here we predict several new effects that arise after a second quench, mostly associated with the extent to which crowding can slow down `fractionation', i.e.\ equilibration of compositions. They include long-lived regular arrangements of secondary domains; wetting of fractionated interfaces by oppositely fractionated layers; `surface'-directed spinodal `waves' propagating from primary interfaces; a `dead zone' where no phase separation occurs; and, in the case of three-phase coexistence, filamentous morphologies arising out of secondary domains.} \\

\end{tabular}

 \end{@twocolumnfalse} \vspace{0.6cm}

  ]

\renewcommand*\rmdefault{bch}\normalfont\upshape
\rmfamily
\section*{}
\vspace{-1cm}


\footnotetext{\textit{$^{a}$Disordered Systems Group, Department of Mathematics, King's College London, London, United Kingdom. E-mail: pabloodecastro@gmail.com}}
\footnotetext{\textit{$^{b}$Currently at Departamento de F\'{i}sica, Facultad de Ciencias F\'{i}sicas y Matem\'{a}ticas, Universidad de Chile, Avenida Blanco Encalada 2008, Santiago, Chile}}
\footnotetext{\textit{$^{c}$Institut f\"ur Theoretische Physik, Georg-August-Universit\"at G\"ottingen, 37077 G\"ottingen, Germany}}



\section{Introduction}
\label{IntroDQ}

The constituent particles of soft matter systems typically exhibit variation in terms of some attribute such as their size, charge, etc.\cite{poon2002physics,van2009experimental,stuart1980polydispersity,evans1998universal,PhysRevE.77.011501} Examples of these multi-component fluid mixtures include colloids, liquid crystals, and polymers.\cite{vieville2011polydispersity,auer2001suppression,belli2011polydispersity,sollich2005nematic,ROGOSIC19961337} 
Although understanding the physical consequences of polydispersity is a considerable challenge, it is has been largely recognised as a matter of both practical and fundamental relevance.\cite{PeterReview,PabloPeter2,decastro2019,zhang2019temperature} In ref.\ \citenum{PabloPeter1}, we considered generic colloidal fluid mixtures subject to typical gas-liquid phase separation from an initial well-mixed state triggered by a single temperature quench into the two-phase coexistence region. We showed that, due to slow changes in composition, (i) dense colloidal mixtures phase-separate in two stages (in agreement with a conjecture by Warren\cite{A809828J}) and (ii) the denser phase contains long-lived `composition heterogeneities'. The effect is obviously impossible in one-component fluids because in that case the composition is the same everywhere (all particles are of the same kind) and only the total local density fluctuates. Firstly we observed that gas bubbles\textemdash or more generally domains\textemdash can be formed with strongly `fractionated' interfaces, where \textit{fractionation} is generally defined as occurring when particle species are unevenly distributed into a new phase or interface. In particular, for a binary mixture composed of two colloidal species plus a passive solvent the gas-liquid interface can then be much richer in one of the two components, in comparison to the initial well-mixed `parent phase'. We found that when a bubble shrinks into non-existence by means of the Ostwald ripening mechanism, remnants of its interface survive for a long time as they struggle---due to crowding---to equilibrate their compositions with that of the surrounding liquid.

In the present work we explore a range of novel manifestations of composition heterogeneities in colloidal mixtures. As in ref.\ \citenum{PabloPeter1}, these effects are driven by crowding but here we study phase separation after a two-step or `double' temperature quench. Specifically we consider phase separation after a second, deeper quench following an initial two-phase separation triggered by a first quench. In this scenario, one expects that secondary domains are transiently formed within phases, before eventually being reabsorbed into the larger domains. We place particular emphasis on low-temperature or `deep' second quenches. In this regime we find interesting effects primarily for dense systems but also more broadly.

The study of the nonlinear time evolution of phase-separating structures in double quench processes can be viewed as part of the larger endeavour of seeking to understand history dependence in phase separation,\cite{CatesRamp} and accordingly has seen increasing interest in recent years. One of the motivations is the fact that it can lead to strategies for controlling structure formation,\cite{PuriControl} in particular with regards to tailoring domain formation in polymeric materials.\cite{Lamellar} Also, the microstructure of food, paint, biological fluids, etc.\ is determined not only by their constituents but also by an (often arrested) process of structural evolution that depends on external conditions such as temperature
.\cite{PuriControl} Double quenches have been explored numerically\cite{Holyst2002,Podariu2007,PuriControl,Lamellar,ClarkeTwoStep,WangLi2011ViscousTwoStep} and experimentally,\cite{HashimotoExperimental1,HashimotoExperimental2,TanakaPRLExperimental,Tanaka2004Experimental} but on the theoretical side only a handful of studies can be applied to colloidal mixtures, none of them looking at effects of slow composition changes or even at compressible fluids. In ref.\ \citenum{ClarkeTwoStep}, the double quench is modeled via the Cahn\textendash Hilliard equation for incompressible symmetric binary polymer mixtures, using a Flory\textendash Huggins free energy functional. After a first quench into the unstable region, the mixture phase-separates via spinodal decomposition into domains that subsequently coarsen; a second quench is then applied to take the system to a point further inside the unstable two-phase region. A new structure factor peak arises from this second quench and eventually disappears as the system progresses towards a single-domain final equilibrium state.

Here we consider `compressible' binary colloidal mixtures subject to secondary temperature quenches into both two- and three-phase regions. The results will prompt us to revisit single quenches but now starting the computational experiment from an already-formed slab domain of well-mixed fluid mixture surrounded by a bath of very dilute vapour. This is similar to one of the geometries considered in a model of binary alloys in ref.\ \citenum{PhysRevLett.78.4970} and \citenum{Plapp99}, where phase separation starts first at the interface between the vapour and the slab. In our case we will see that, within the slab domain, other domains (of gas, i.e.\ bubbles) are formed transiently in a way that can be connected to the double quench experiment. This is because both situations can be viewed as instances of `domain-within-domain' (or `secondary') phase separation. The observed effects described below include long-lived regular arrangements of secondary domains; interrupted coarsening of primary domains; wetting of fractionated interfaces by oppositely fractionated layers; `surface'-directed spinodal `waves' propagating from primary domain interfaces; a `dead zone' where no phase separation occurs; and, in the case of a second quench into the three-phase region, filamentous morphologies arising out of secondary domains.

The Polydisperse Lattice-Gas (PLG) mean-field kinetic theory derived in ref.\ \citenum{PabloPeter1} will be employed in this work in both analytical and numerical approaches. (We also calculate the equilibrium compositions as we did in ref.\ \citenum{PabloPeter1}, i.e.\ by solving the coexistence conditions.) As we will recall below, the theory includes two elementary processes: particle-solvent and particle-particle exchanges. Their relative rate will be used to tune and characterise the effects of slow composition changes in the form of composition heterogeneities. For the sake of simplicity only mixtures of two attractive species plus `vacancies' (or passive solvent particles) will be considered, but we expect the phenomena observed here to generalise to a broad class of polydisperse systems as shown and explored in ref.\ \citenum{PabloPeter1}.

The remainder of this paper is organised as follows. In Section \ref{KineticPLG} we review the basic equations derived in our model in ref.\ \citenum{PabloPeter1}, which form the basis of the subsequent analyses. In Section \ref{DeepQuench} we study slow-fractionation effects after secondary quenches into the two-phase region. Then in Section \ref{SlabDomain} we consider the slab domain as initial state. In Section \ref{threephase} we discuss our simulations for secondary quenches into the three-phase region. Section \ref{ConclusionsDouble} concludes the paper with a summary, discussion and pointers towards future work.

\section{Kinetic PLG model}
\label{KineticPLG}
We recall briefly the PLG model\cite{PhysRevE.77.011501} of polydisperse colloidal systems, as well as its kinetic version,\cite{PabloPeter1} which has been used to generate the simulations in this work. The model is described by the Hamiltonian
\begin{equation}
H = - \sum\limits_{\langle i,j\rangle}^{}\sum\limits_{\alpha, \beta}^{}\sigma_{\alpha}\sigma_{\beta}n_{i}^{\alpha}n_{j}^{\beta}
\label{plg}
\end{equation}
where $i$ runs over the sites of a periodic lattice $i =1, \dots ,L^{D}$, assumed simple cubic and $D$-dimensional in this work, with lattice spacing $a=1$, unless otherwise stated; the sum runs over all pairs $\langle i,j\rangle$ of nearest-neighbour sites; $\sigma_{\alpha}$ is the value of the polydisperse attribute associated with colloidal species $\alpha$, which controls the strength of interparticle interactions; it is a positive number for attractive interactions as considered here. Generically one can consider a mixture with $M$ species, with the summations over $\alpha$ and $\beta$ therefore running from $1$ to $M$. In the next sections we will use $M=2$. The (occupation) variable $n_{i}^{\alpha}$ simply counts the number of particles of species $\alpha$ at site $i$, so that $\alpha$ is not an exponent, but a superscript. A hard-core constraint is imposed:
\begin{equation}
\sum\limits_{\gamma=0}^{M} n_{i}^{\gamma}=1\quad \forall i
\label{hardcore}
\end{equation}
where $n_i^{0}$ refers to \textit{vacancy}, i.e.\ $n_i^{0}=1$ indicates the presence of a vacancy at site $i$, or, equivalently, a solvent particle. Observe that the solvent particles are `passive'\cite{ignacio} in this framework, in such a way that any non-hydrodynamic effect caused by them has already been included in the effective interaction between the particles as described by eqn (\ref{plg}). (Hydrodynamic interactions mediated by the solvent are ignored.) We therefore refer to the solvent particles as `vacancies' in the following. We will use the letter $\gamma$ as the species index for summations running from $0$ to $M$, while for summations running from $1$ to $M$, we use $\alpha$ (or $\beta$), unless otherwise specified. The overall number density of particles of species $\alpha$ is a number between $0$ and $1$ and is denoted by $p^{\alpha}$. By summing up $p^\alpha$ over $\alpha=1,\dots,M$ one obtains the overall total density, denoted by $\rho$; when added to the number density of vacancies this gives $1$.

In ref.\ \citenum{PabloPeter1} a kinetic version of the PLG model was developed. Defining a time-dependent local density for each species as $p_i^{\alpha}(t)\equiv\langle n_i^{\alpha} \rangle (t)$, where $\langle\dots\rangle$ denotes configurational average, we obtained the mean-field kinetic equations
\begin{equation}
\begin{aligned}
\frac{dp_i^{\alpha}}{dt}=-\sum\limits_{j\in\partial i}^{}\sum\limits_{\gamma=0}^{M}\left[ \frac{p_i^{\alpha} p_j^{\gamma}	w^{\alpha\gamma}}{1+\exp{\left(\bigl\langle\Delta H_{ij}^{\alpha\gamma}\bigr\rangle/T\right)}}- \frac{p_j^{\alpha}p_i^{\gamma}	w^{\alpha\gamma}}{1+\exp{\left(\bigl\langle\Delta H_{ji}^{\alpha\gamma}\bigr\rangle/T\right)}}\right]
\end{aligned}
\label{kineticequations}
\end{equation}
valid for any site $i$ and species $\alpha$, where $T$ is the temperature and -- within the mean-field approximation -- one has
\begin{equation}
\begin{aligned}
\bigl\langle\Delta H_{ij}^{\alpha\gamma}\bigr\rangle ={}
& \sum\limits_{\beta}^{}\left(\sum\limits_{k\in \partial i}^{}\epsilon_{\alpha\beta}p_{k}^{\beta}-\sum\limits_{l\in \partial j}^{}\epsilon_{\alpha\beta}p_{l}^{\beta}\right)\\ 
&-\sum\limits_{\beta}^{}\left(\sum\limits_{k\in \partial i}^{}\epsilon_{\gamma\beta}p_{k}^{\beta}-\sum\limits_{l\in \partial j}^{}\epsilon_{\gamma\beta}p_{l}^{\beta}\right)
\end{aligned}
\label{deltaHMF}
\end{equation}
where $\epsilon_{\alpha\beta}\equiv\sigma_{\alpha}\sigma_{\beta}$, and we have introduced the notation $j\in\partial i$, meaning that the summation has to be performed over all nearest-neighbour sites $j$ of
site $i$. Also, $w_{i j}^{\alpha\gamma}$ is the \textit{jump rate} for an $\alpha$-particle at site $i$ to exchange positions with a $\gamma$-particle at site $j$ (or with a vacancy, if $\gamma=0$). Effectively, we are considering a kinetic model equipped with two elementary processes: (i) the jump of a particle from an occupied lattice site to an empty one, and also (ii) the direct interchange between two particles from arbitrary species. Notice that the overall number of particles of each species remains constant. Although the physical elementary processes are the jumps to empty sites, i.e.\ particle-vacancy exchanges, for the subsequent analysis it is useful to include also direct particle-particle swaps. 
Here, we use Glauber-like jump rates:
\begin{equation}
w_{i j}^{\alpha\gamma}=w^{\alpha\gamma}\left[\frac{1}{1+\exp{\left(\langle\Delta H_{ij}^{\alpha\gamma}\rangle/T\right)}}\right]
\end{equation}
where $\Delta H_{ij}^{\alpha\gamma}$ is the energy difference associated with the jump; note that the configurations after and before the jump are identical except for the exchange between $\alpha$ and $\gamma$. The prefactor $w^{\alpha\gamma}$ is an `attempt rate'. This gives the actual jump rate $w_{i j}^{\alpha\gamma}$ when $\langle\Delta H_{ij}^{\alpha\gamma}\rangle$ is large and negative, while it is reduced exponentially for large positive energy changes. Hereafter we will set $w^{\alpha0}=w_0$ and $w^{\alpha\beta}=w_s$, for any $\alpha\neq0$ and $\beta\neq0$, where $w_0$ and $w_s$ are constant attempt rates associated with particle-vacancy and particle-particle exchanges, respectively. (The `$0$' subscript is for vacancy, and the `$s$' is for \textit{swapping}.)

\subsection{Warren's scenario}
As indicated above, crowding can lead to a two-stage scenario\cite{A809828J} that is only possible in mixtures. This idea by Warren can be summarised as follows. Start by noticing that a phase-separating mixture will need to locally change composition, apart from changing its local total density. This is needed in order to achieve equilibrium fractionation. Thus, in dense systems, particles of different species are expected to push against each other as the system progresses towards the correct species balance, in a process that effectively requires particles of different species to be exchanged. This results in a slowdown of composition equilibration. Therefore fractionation might be quantitatively insignificant for a long initial period, whereas total density equilibration would occur much earlier as it can proceed via motion of particles of \textit{any} species. One would then expect the dynamics to be initially dictated by a phase diagram that `artificially' prohibits fractionation. (This is called the \textit{quenched} phase diagram.) Only over much longer timescales would the system start to decrease its free energy further by creating fully fractionated domains, which at that stage will conform to the actual equilibrium phase diagram (dubbed the \textit{annealed} phase diagram). As one can see the dynamics has two stages. 

In the late-time regime Warren's scenario can also manifest itself, now in its general form, which simply says that any equilibration of compositions will be slow in dense systems. An example of this are the long-lived composition heterogeneities arising when domains with highly fractionated interfaces shrink and leave behind regions whose composition struggles to equilibrate. We have provided several pieces of evidence for the above ideas in ref.\ \citenum{PabloPeter1}. A way to destroy Warren's scenario is to allow the exchange between particles of different species in the kinetic PLG model described above, i.e.\ by turning on $w_s$, as this facilitates fractionation. This will be explored below as a way of detecting which dynamical effects (in the conventional dynamics with $w_s=0$) are due specifically to the slow nature of particle exchanges.

\section{Deep secondary quench}
\label{DeepQuench}

We study two-step temperature quenches where the final, second temperature quench takes us further into the unstable region. Phase separation is initiated only via spinodal decomposition within our approach as to capture `nucleation and growth' events one would need to introduce noise to
the deterministic mean-field kinetic eqns (\ref{kineticequations}). We will refer to the structures formed after the first quench as primary domains. These will be allowed to coarsen before we perform a secondary quench, which in turn may or may not give rise to the formation of secondary domains within each of the primary domains. We will be interested primarily in investigating the secondary phase separation within the primary liquid phase rather than in the gas. This is because the liquid is at higher density and so much more likely to show the effects of slow composition changes. In line with this focus, we choose the secondary quench temperature low enough to make the primary liquid phase unstable. If we let the primary phase separation run for long enough to obtain nearly equilibrated primary phase compositions, we can say that the liquid will be unstable following the secondary quench if the second quench temperature satisfies
\begin{equation}
T\leq z\left(\rho_2-\rho_1^2\right)
\label{InstabilityInequality}
\end{equation}
where the right-hand side is the (`annealed') spinodal temperature given by eqn (6) from ref.\ \citenum{PabloPeter1} and is evaluated at the composition of the equilibrium liquid obtained from the primary phase separation; here we made use of the moment densities $\rho_{n}\equiv\sum\limits_{\alpha}^{}\sigma_\alpha^{n}  p^{\alpha}$ and the lattice coordination number $z$. The occurrence or not of an instability in the primary gas phase can be calculated analogously. Moreover, the minimal `size' of the secondary structures must be smaller than the length scale of the primary domains. We quantify this by the secondary phase separation spinodal length, i.e.\ $2\pi/k_{\rm max}$, where $k_{\rm max}$ is the wavenumber of the fastest growing fluctuation mode.
(See ref.\ \citenum{PabloPeter1} and eqn (24) therein.)

In order to access clearly distinct behaviours---slow versus fast composition changes---we will compare simulations performed for particle-particle swapping rates $w_s=0$ and $w_s=1$ at deep secondary quenches but remaining for now within the two-phase region. The primary liquid phase is both unstable and dense enough to trigger slow-fractionation effects in the case of $w_s=0$. As in ref. \citenum{PabloPeter1}, we will start from a parent system with equal amounts of the two colloidal species $A$ and $B$. The normalized composition, therefore, is $(0.5,0.5)$ and the polydisperse attributes are $\sigma_A=1+d$ and $\sigma_B=1-d$, where $d$ is chosen between $0$ and $1$, being a measure of the ``polydispersity'' of the fluid. The colour scheme in the simulation snapshots throughout the paper is such that dense regions with a high concentration of $A$'s are blue, those with predominantly $B$'s are red and dilute regions dominated by vacancies are white.\bibnote{More specifically, to determine the colour of a site $i$, we blended together the colours red, green, and blue. The intensity of each of these colours at a given site varies from $0$ to $1$. In our scheme, red, green, and blue intensities are given by $1-p_i^A,p_i^0,1-p_i^B$, respectively. (Remember the notation for the local concentration of vacancies, i.e.\ $p_i^0=1-p_i^A-p_i^B$.) This leads to the colour key shown in the top-left part of Fig.\ 8 in ref.\ \citenum{PabloPeter1}. It is plotted in triangular colour space in $(p^A, p^B, 1 - p^A - p^B)$, dropping the site index $i$. For example, if the concentration of particles of species $A$ at one site is high (low), and the concentration of particles of species $B$ at the same site is low (high), then the site colour will tend towards blue (red); if the concentrations of all species are all low, then the site colour will tend to white.} The simulations are performed in a
square lattice in $D = 2$, i.e.\ $z = 4$. In the low-$T$ that we mainly focus on, the final equilibrium liquid has very little fractionation since the gas is extremely dilute and therefore the vast majority of the particles are in the liquid. Two caveats arise in using such low temperatures: we get a strong slowdown, with a consequent increase in computational times; also, we see some lattice effects such as `square' interfaces. This happens due to the system's difficulty in overcoming energy barriers that prevent the formation of more rounded domains. The former issue does not pose a significant problem since simulations of the final asymptotic coarsening where the secondary domains have disappeared will be beyond our scope. The latter would not be expected to affect the physics qualitatively, though this is something that would be worth checking in future work via e.g.\ off-lattice simulations. 

\begin{figure}[h]
	\centering
	\includegraphics[width=\columnwidth]{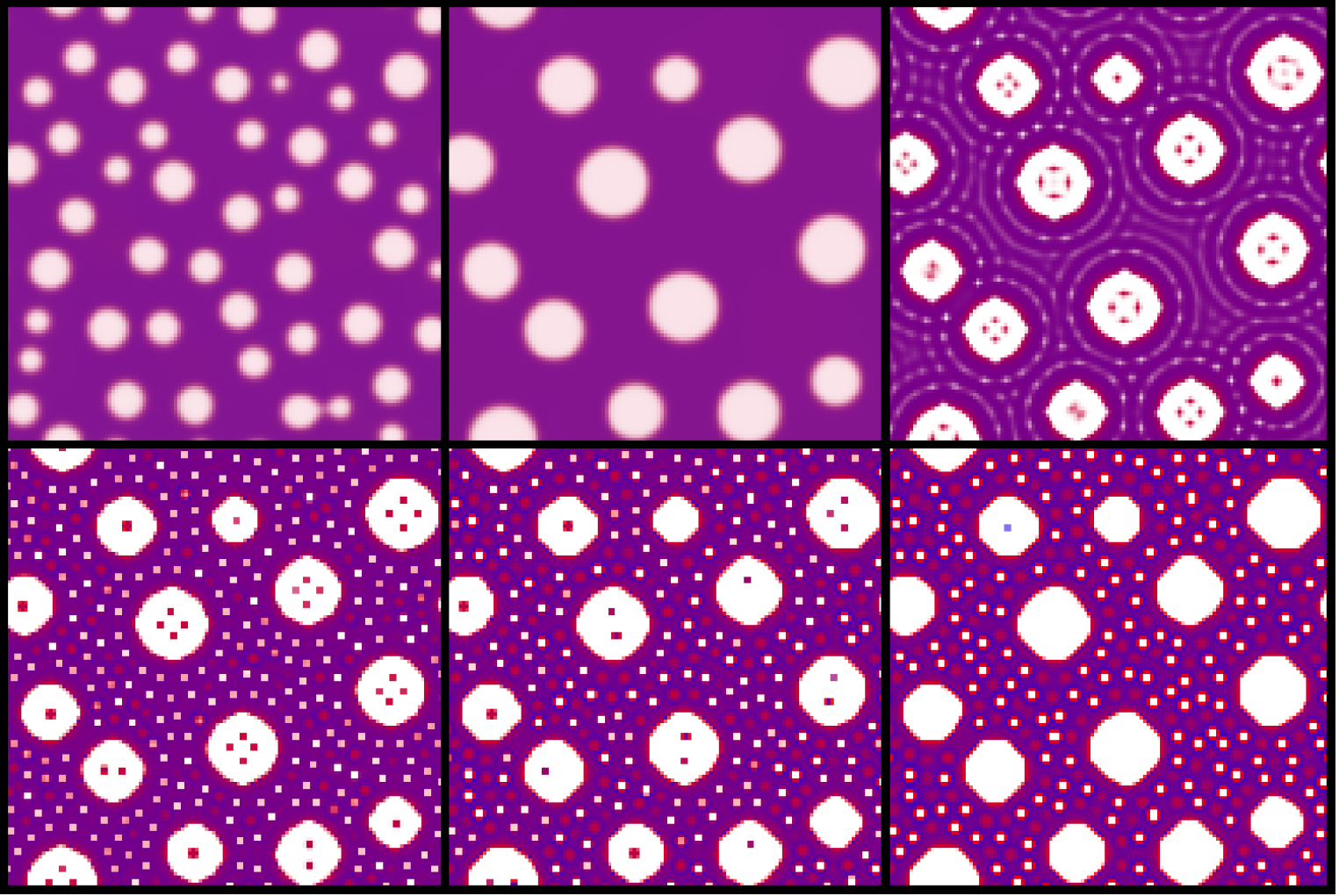}
	\caption{Time snapshots showing the local compositions throughout the system, as from numerics. For the colour scheme see main text. The lighter portions of the system are gas bubbles, which are surrounded by a $B$-rich interface separating them from a liquid phase. Parameters: $p^A=p^B=0.375$, $T_{1}=0.7$, $T_2=0.1$, $d=0.15$, $w_s=0$, and $L=128$. From top left to bottom right, the snapshots are taken at $t=4717$, $40000$, $40009$, $50705$, $235000$, and $3.3757 \times 10^6$. The secondary quench was performed at $t=t_2=40000$. The symmetries are a consequence of the lattice structure and of the low final temperature.}
	\label{TwoStepWs0}
\end{figure}

Fig.\ \ref{TwoStepWs0} shows snapshots at times both before and after a secondary quench into the two-phase region of a binary mixture. (For the simulations shown in this paper we have found it useful to implement an adaptive time-stepping method like the one studied in ref.\ \citenum{li2017computationally}.) The system's overall density of colloidal particles is $0.75$. The first quench, to a temperature $T=T_1=0.7$, is shallow and generates a primary liquid that is only slightly fractionated and has a high overall density of $0.92$. With respect to the $\rho$-$T$ phase diagram, the first quench places the system at a point that is close to the high-density branch of the quenched spinodal curve, to the left of it; because of the relatively small value for $d=0.15$ used here, annealed and quenched spinodal curves lie almost on top of each other but the quenched one still lies entirely inside the annealed spinodal region. The second (deeper) quench down to temperature $T=T_2=0.1$ was then performed at $t=t_2=40000$, a time at which the primary domains have grown substantially and nearly equilibrated their compositions. After some time we can see in the snapshots that small secondary domains have formed. The primary and secondary domains of interest to us are the `large' and `small' gas bubbles, respectively, in the surrounding liquid; tiny liquid droplets also form within the primary gas bubbles but these equilibrate quickly by being reabsorbed through the interface as a result of the fast dynamics at the very low density of the gas phase. The initial lengthscale of the secondary structures formed in the primary liquid is consistent with the spinodal lengthscale predicted theoretically; note that because of the large value of $t_2$ used, the primary bubbles have become sufficiently separated from each other to fit a few such spinodal lengths of the secondary phase separation in between them. 
Note that the secondary spinodal dynamics manifests initially as `waves' propagating from the primary bubbles interfaces, a behaviour that we will discuss further below. (See Section \ref{SlabDomain}.) Out of these waves, the secondary gas bubbles then form. Once these eventually disappear again by shrinking, their fractionated interfaces remain visible as strongly long-lived patches of composition heterogeneities. This is analogous to what we found in ref.\ \citenum{PabloPeter1} but here we can see these composition heterogeneities during time intervals at least two orders of magnitude longer, as a result of the low temperature.

\begin{figure}[h]
	\centering
	\includegraphics[width=\columnwidth]{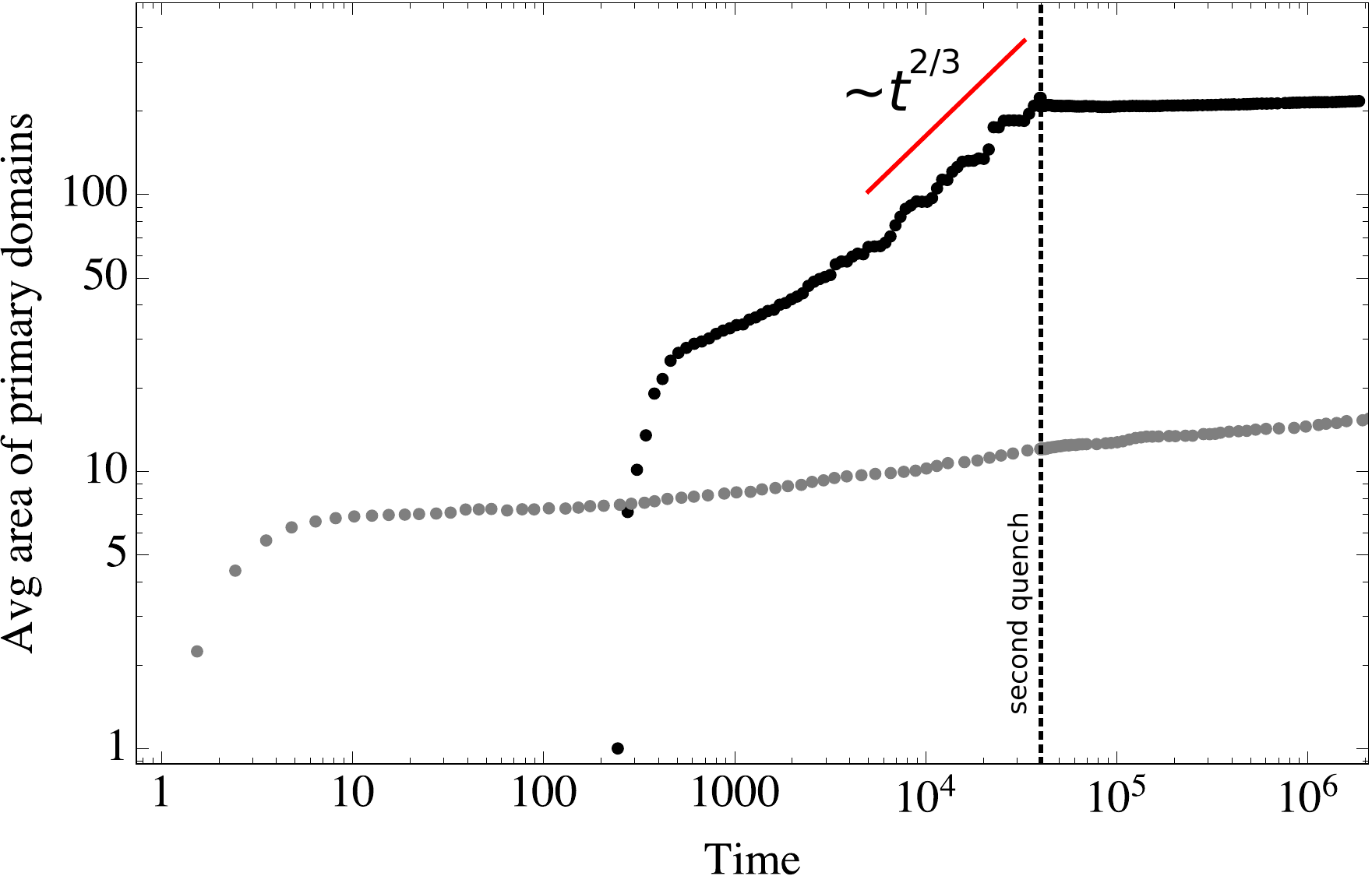}
	\caption{Log-log plot of the average area (in number of lattice sites) of primary domains versus time. The red solid line segment is proportional to $t^{2/3}$. The upper data come from the numerics shown in Fig.\ \ref{TwoStepWs0}, where $T_1=0.7$ and $T_2=0.1$. They show a kink at the time of the second quench before entering a plateau of virtually constant area. The lower data are for a situation where there is no secondary temperature quench and we start directly with a quench to $T_1=T_2=0.1$.}
	\label{Coarsening}
\end{figure}

In Fig.\ \ref{Coarsening} we can see the time evolution for the average area of primary domains. The areas have been calculated via image analysis.\bibnote{This has been performed using a function from \textit{Mathematica 11.2} called `ComponentMeasurements', which has been applied to a binary image obtained from our data; to construct the binary image we defined that total densities below $0.5$ constitute gas bubbles. Therefore the interfaces are mostly considered part of the liquid.} They are given in number of lattice sites enclosed by the domain's interface. In order to distinguish between primary and secondary domains we used the fact that, with our simulation parameters, the largest secondary domain always remained smaller than the smallest primary domain at $t_2$, the time of the second quench.\bibnote{For the $w_s=1$ data, in order for this to be true we needed to use a slightly smaller region rather than the entire system.} According to the Lifshitz-Slyozov (LS) law\cite{LIFSHITZ196135,Huse1986,STREITENBERGER20135026} the average linear domain size for coarsening with a conserved order parameter as here should increase as $l(t)\sim t^{1/3}$ in the asymptotic regime, a behaviour that can be identified in the upper data in Fig.\ \ref{Coarsening}. Since we plot the area rather than linear size the graph shows exponent $2/3$. The asymptotic behaviour is reached during the final stages of the primary phase separation for $t<t_2$. Then the deep quench generates a plateau in the `coarsening' dynamics, so that effectively it interrupts the coarsening of the primary domains. This interruption extends across essentially all of our simulation time after $t_2$, which is qualitatively consistent with the depth of the quench causing a drastic slowdown of the dynamics. Within the plateau in the plot that corresponds to the interrupted coarsening there is in fact a very slow growth because the primary bubbles still absorb gas from the tiny secondary bubbles as they disappear, but this is quantitatively a rather weak effect. Indeed the number of primary bubbles remains constant for all $t\geq t_2$ within our simulation time window, showing that primary coarsening is essentially halted during this period.

\begin{figure}[h]
	\centering
	\includegraphics[width=\columnwidth]{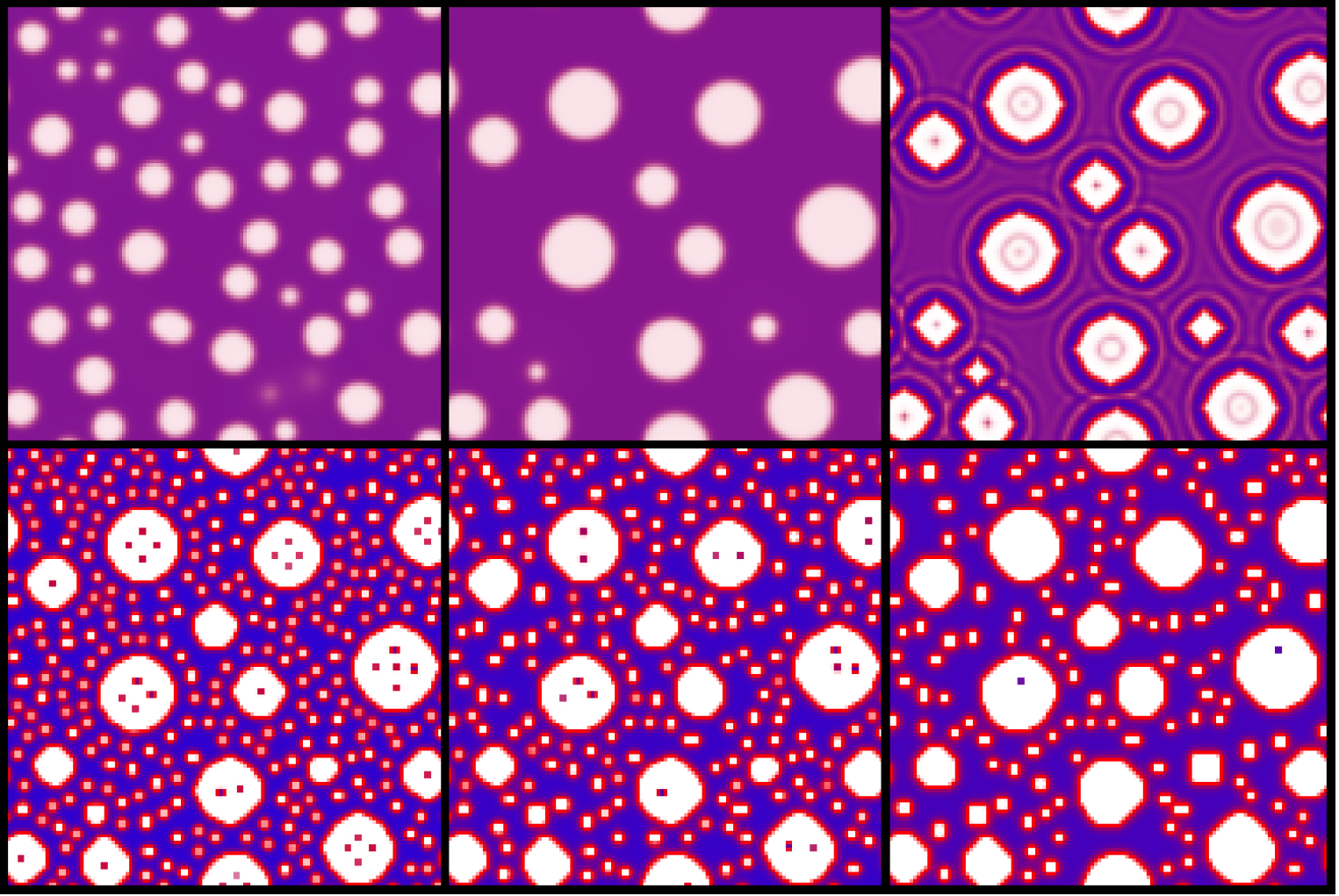}
	\caption{As Fig.\ \ref{TwoStepWs0} but for $w_s=1$. Parameters: $p^A=p^B=0.375$, $T_{1}=0.7$, $T_2=0.1$, $d=0.15$, $w_s=1$, and $L=128$. From top left to bottom right, the snapshots are taken at $t=4256$, $40000$, $40003$, $46220$, $179285$, and $3.3811 \times 10^6$. The secondary quench was performed at $t=t_2=40000$.}
	\label{TwoStepWs1}
\end{figure}

\begin{figure}[h]
	\centering
	\includegraphics[width=\columnwidth]{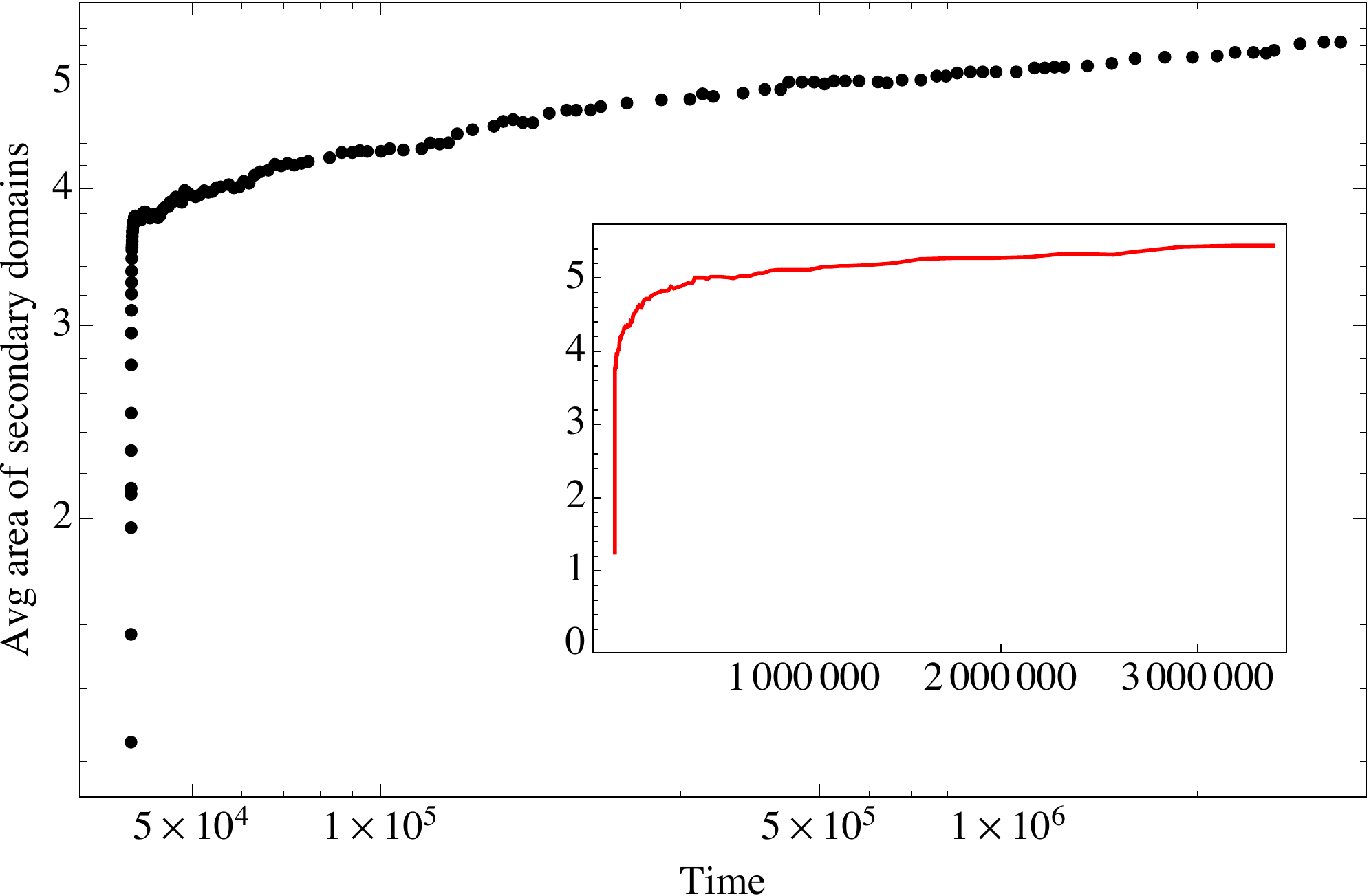}
	\caption{Log-log plot of the average area (in number of lattice sites) of secondary domains. The data used is the same as in Fig.\ \ref{TwoStepWs1}, where $T_1=0.7$ and $T_2=0.1$. Inset: linear scale, with points joined by lines to guide the eye.}
	\label{SecondaryGrowth}
\end{figure}

Fig.\ \ref{TwoStepWs1} shows snapshots for the same parameters, except that now we have $w_s=1$. This effectively `switches on' direct particle-particle swaps and so facilitates composition changes, leading to different morphological features. The secondary bubbles merge by collapsing one into the other, with the smallest ones collapsing first. (By ``collapse'' we mean that one bubble disappears by shrinking, with most of its gas being taken up by a neighbouring bubble.) Initially, the number of secondary bubbles increases very quickly from zero (as in typical spinodal decomposition away from the critical point) and then we have a long period of decrease. The rate of decrease slows over a long period (data not shown) as the remaining bubbles are no longer very close to each other and therefore cannot transfer gas so easily.
In this process the interface material of the secondary bubbles that disappear becomes partly dispersed into the liquid and partly used to increase the interface lengths of surviving bubbles, rather than becoming trapped as long-lived composition heterogeneities. For the average area of these secondary domains, shown in Fig.\ \ref{SecondaryGrowth}, we initially have a strong increase; subsequently, since bubbles become increasingly farther apart, we have that despite the faster dynamics ($w_s=1$) their average area increases only slowly for a very long period. The slow rate of increase is partly due to the low temperature. In addition, the presence of the large primary bubbles with their low Laplace pressure counteracts the tendency of secondary bubbles to coarsen. Rather than reaching an asymptotic $t^{1/3}$ coarsening regime, the secondary bubbles are then expected to eventually shrink and disappear before coarsening of the primary bubbles resumes.

By comparison, for $w_s=0$ we have that, following a similar long transient where the average area of secondary bubbles increases---and their number decreases---, a much longer period of essentially arrested behaviour occurs; this period starts beyond the penultimate snapshot in Fig.\ \ref{TwoStepWs0} and lasts until the end of our simulation time window. (Data not shown.) We hypothesize that the slow kinetics of composition changes is again at play here. For example, it is more difficult for the system to increase the length of interfaces of secondary bubbles since interface material from neighbouring shrunk bubbles becomes trapped as long-lived heterogeneities. This will favour gas liberated by the disappearance of secondary bubbles to go into the primary bubbles, where it can be taken up with a smaller increase in interfacial length.

Interestingly, the evolution of the average area of the primary domains for $w_s=1$ remains qualitatively similar to that for $w_s=0$ in Fig.\ \ref{Coarsening} (data not shown). The implication is that while secondary bubbles do collapse for $w_s=1$, the amount of gas that such processes transport into primary bubbles -- rather than other secondary bubbles -- remains small here. 

After the coarsening-interruption plateau, we expect for both $w_s=0$ and $w_s=1$ that at very late times the system will eventually resume coarsening of the primary bubbles, following a LS-law but with a much reduced prefactor due to the significantly smaller temperature. By this time all secondary bubbles will have disappeared. 

To probe this issue further, we show the lower data in Fig.\ \ref{Coarsening}, which corresponds to a situation where we start directly at the low temperature $T_1=T_2=0.1$, i.e.\ we perform a single quench. By the end of our simulation window this phase separation has not yet reached the LS-law coarsening regime as the bubbles are small and grow very slowly. Now, at sufficiently late times the low-$T$ coarsening dynamics is expected to lose all memory of having started at higher $T$. Therefore the single-quench data should connect smoothly with the final coarsening curve of the two-step case. But even if the single-quench case reached the LS-law regime soon after the end of our simulation window, its prefactor shows that this would cross the two-step plateau only at a much longer time. This means that by the end of our double quench simulation window we are still far from the late-time edge of the coarsening-interruption plateau. This conclusion is valid for both $w_s=0$ and $w_s=1$.

\begin{figure}[h]
	\centering
	\includegraphics[width=\columnwidth]{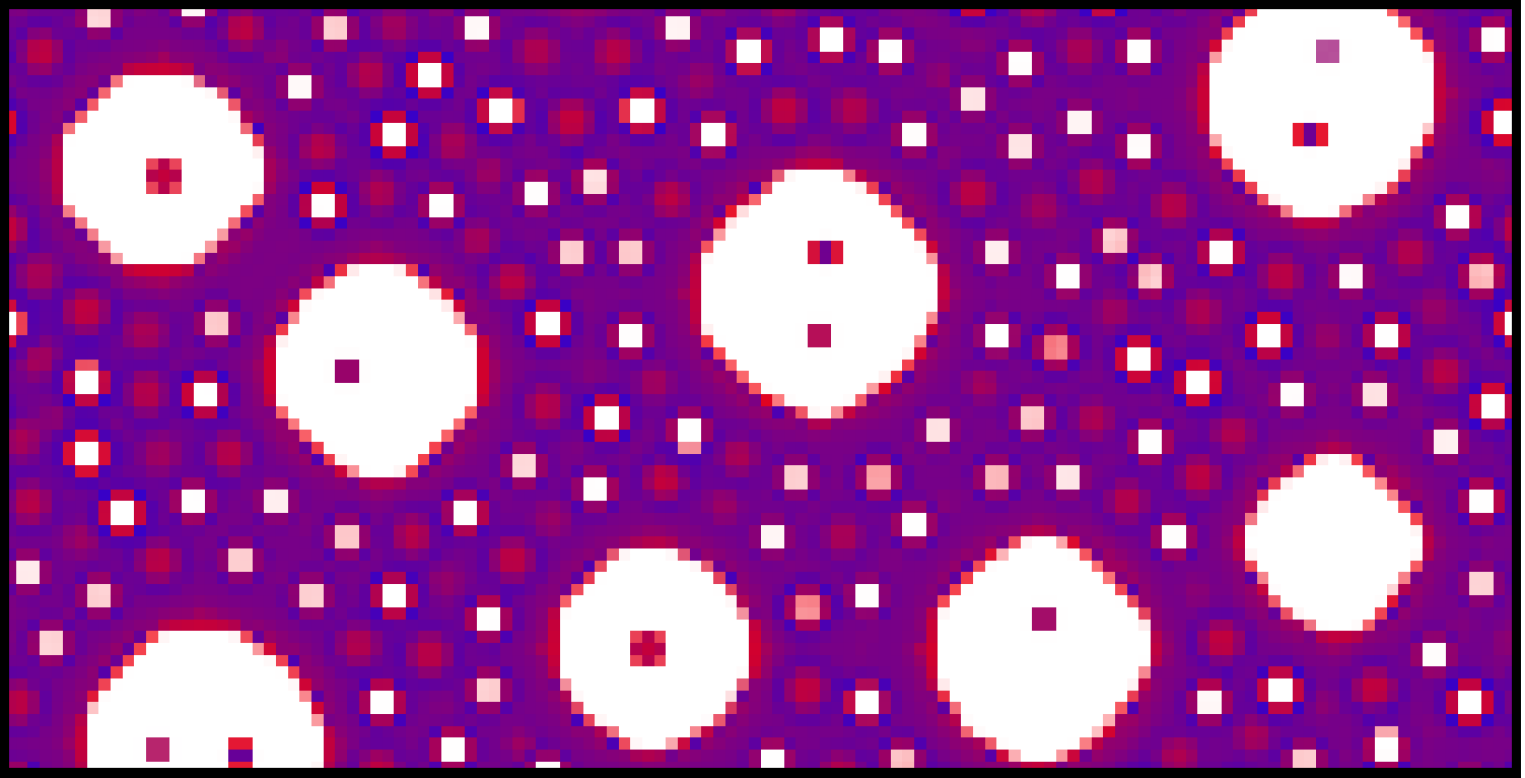}
	\caption{Lower half of bottom-centre snapshot ($t=235000$) in Fig.\ \ref{TwoStepWs0}.}
	\label{LowerHalf}
\end{figure}

Some other distinctions between the $w_s=0$ and $w_s=1$ cases can be seen by examining the snapshots in Figs.\ \ref{TwoStepWs0} and \ref{TwoStepWs1} closely. First, note that primary bubbles that are very close together might not leave enough space for a spinodal length to fit in between them. This effective confinement can prevent the spinodal growth of fluctuations and suppress secondary phase separation in the relevant regions of the system. Although rather difficult to discern in the numerical example for $w_s=0$ in Fig.\ \ref{TwoStepWs0} (but easier to see in the enlarged representation in Fig.\ \ref{LowerHalf}), this is what appears to happen for instance with the closest pair of primary bubbles on the left-hand side of the system: notice how the composition of the environment between those bubbles at $t=235000$ is more similar to that of the primary liquid, consequently forming a very subtle `bridge' that wets the two bubbles. This kinetic effect eventually disappears as composition equilibration progresses; in the fast dynamics for the $w_s=1$ case it simply does not exist.

The liquid phase composition evolves in interestingly different ways indeed, depending on $w_s$. We point out the different morphologies directly after the second quench. For $w_s=1$ the spinodal waves quickly lead to the formation of a substantially $A$-rich liquid phase and do so in a largely homogeneous manner. The enrichment in $A$-particles compensates for the existence of a large amount of strongly $B$-rich interfaces. In fact note that the formation of interfaces generally depletes $B$-particles from the liquid next to the interface. For the case $w_s=0$ the composition changes involved in this depletion process are too slow to cause significant effects and therefore the spinodal waves lead to a liquid without much fractionation, as one can see from the top-right snapshot (directly after $t=t_2$) in Fig.\ \ref{TwoStepWs0}. 

At later times (for $w_s=0$) the secondary bubbles grow slightly, at the expense of smaller ones, and of course need vacancy input to do so. (In the Sigehuzi-Tanaka experiment in ref.\ \citenum{Tanaka2004Experimental} the growth of the secondary domains is also extremely slow.) As vacancies get into the bubbles we temporarily have regions of smaller density surrounding these domains. This allows for composition changes to happen faster than in the bulk liquid. Thus this contributes to the formation of $A$-rich regions surrounding the growing secondary bubbles while the rest of the system remains relatively poorly fractionated. (See Fig.\ \ref{LowerHalf}; note that in the regions between secondary bubbles there is depletion of $B$'s from both sides.) We expect this composition-heterogeneous liquid to eventually become homogeneous again---this stage is beyond our simulation window. If it occurs at a point where there is still a large density of strongly $B$-rich interfaces then we should expect to see a state similar to the $A$-rich homogeneous liquid that is formed shortly after the secondary quench in the case ${w_s=1}$. Otherwise, the liquid might never become homogeneously $A$-rich: remember that at equilibrium almost all of the $B$-rich interface material will have dispersed (we have only one domain) and the liquid is, again, only slightly fractionated rather than $A$-rich.

In order to investigate the above and other related phenomena in more detail we will in the next section 
switch to a simpler set-up.

\subsection{Secondary domain coarsening and LSW theory}

Before moving on we will briefly set out a simple theoretical picture for the absence of an asymptotic coarsening regime for the secondary bubbles. A quantitative way to understand this would be within the LSW theory\cite{wagner1961theorie}. This assumes that bubbles are dilute and that chemical potentials equilibrate quickly to the presence of the bubble interfaces, neither of which are obviously true in our case but may still be reasonable approximations. In such a theory the radius of each bubble will change as
\begin{equation}
\frac{dR}{dt} \propto \frac{1}{R} \left(\frac{1}{R_c}-\frac{1}{R}\right)
\label{RadiusChange}
\end{equation}
such that bubbles larger than $R_c$ will grow while smaller ones will shrink.
In eqn (\ref{RadiusChange}), $R_c$ is determined indirectly by the requirement that the domain size density $n(R,t)$ must obey $\int R^D n(R,t) dR = \text{const}$.\ (with dimension $D=2$ in our case; this constraint just expresses the fact that the total gas volume remains the same). Differentiating the constraint with respect to time and inserting (\ref{RadiusChange}) gives
\begin{equation}
R_c = \frac{\int R^{D-2} n(R,t) dR}{\int R^{D-3} n(R,t) dR}
\end{equation}
For $D=2$ specifically we have $R_c = 1/\langle1/R\rangle$ where $\langle\dots\rangle$ is an average over the normalized bubble size distribution. In this average the primary bubbles make little contribution initially because they are much fewer in number and have a larger $R$. As long as this remains the case the secondary bubbles will then coarsen as if the primary bubbles were not present: $R_c$ is set by the secondary bubble population and its value lies well below the typical primary bubble radius.

As the number of secondary bubbles decreases and their size increases, $\langle1/R\rangle$ will pick up the contributions from the primary bubbles---which will still have larger $R$ than the secondary ones---and so become smaller, making $R_c$ larger and thus forcing more of the secondary bubbles to shrink. Eventually, as $R_c$ becomes of the order of the primary bubble radius, all smaller secondary bubbles will shrink and disappear, and coarsening of the primary bubbles (along with a few surviving larger secondary bubbles) will resume.

The above analysis suggests that in principle one \textit{could} see a $t^{1/3}$ coarsening of secondary bubbles during an intermediate time regime, provided the primary bubbles are very large at the time of the second quench. Even in this case, of course, the secondary bubbles will eventually disappear again as their size approaches that of the primary ones.

\section{Phase separation from slab geometry}
\label{SlabDomain}
After this intermezzo we return to our main thrust and consider a liquid domain in the form of a slab surrounded by vapour, and perform a single quench from that state. As anticipated in the introduction, this is similar to one of the geometries considered in a model of binary alloys in ref.\ \citenum{Plapp99}, where phase separation starts first at the slab-vapour interface. Importantly, however, the interaction parameters used in ref.\ \citenum{Plapp99} lead to `A-B' phase separation rather than a fractionated gas-liquid coexistence as in our case. The slab setup also connects closely to the double quenches considered so far, as starting from a slab initial condition is equivalent to what one would obtain by taking $t_2\rightarrow \infty$ in a double quench situation. Only a single domain would be left at the time of the second quench, forming either a slab or a single circular domain in a finite system. (We can therefore also think of the slab domain as the liquid between two big---and therefore locally flat---primary bubble interfaces.) Thus in this section we use the slab domain as our initial state to obtain more detailed insights in a geometry that is simpler to analyse. Initially we will imitate the conditions of the simulations in Figs.\ \ref{TwoStepWs0} and \ref{TwoStepWs1} as a baseline, before proceeding to cases with higher final temperature and $d$ where we will see additional effects.

\subsection{Deep quench `revisited'}
\label{revisited}
\begin{figure}[h]
	\centering
	\includegraphics[width=\columnwidth]{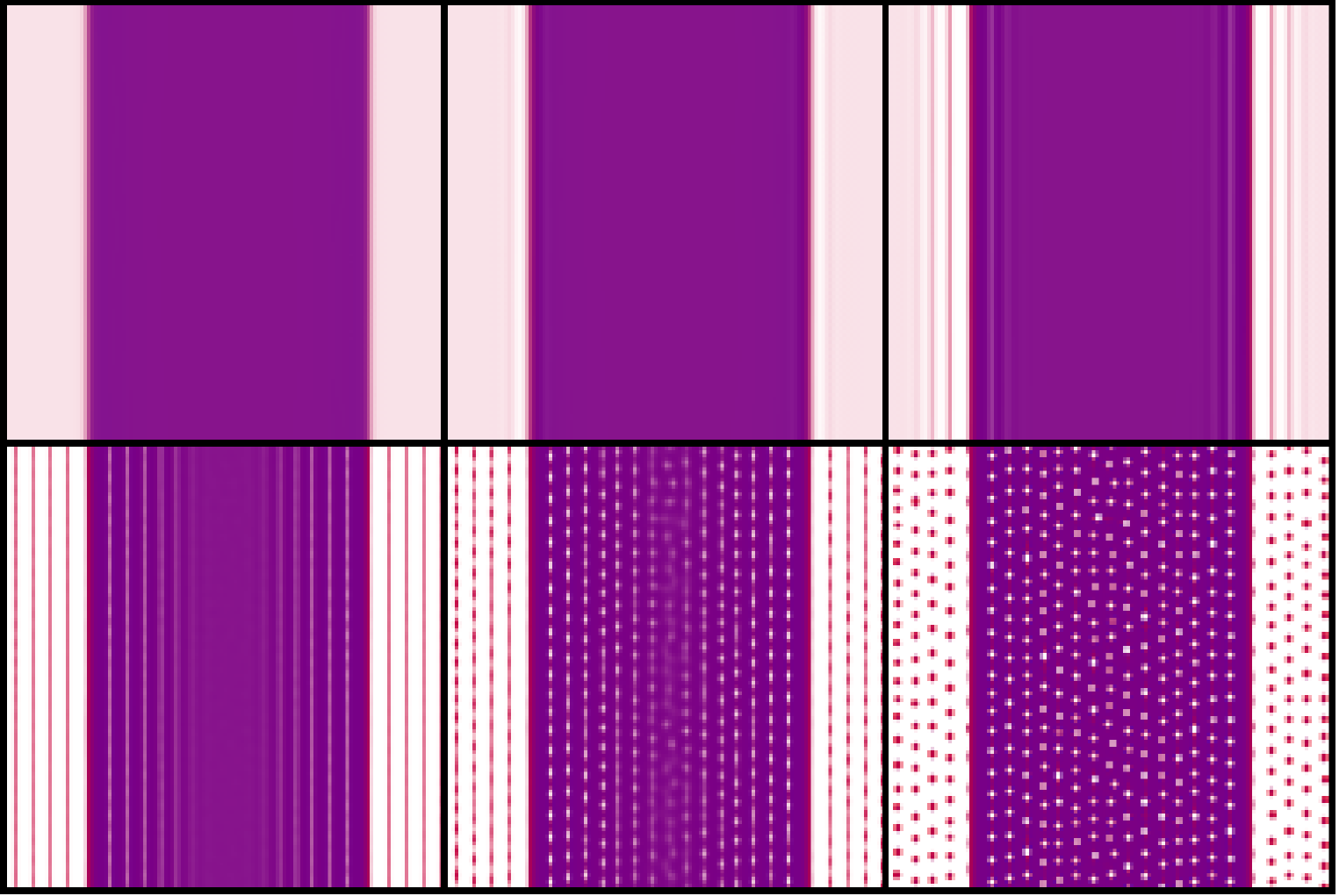}
	\caption{Time snapshots starting from an equilibrated slab-geometry liquid domain surrounded by a vapour phase. The initial liquid and vapour compositions as well as the other parameters match those at the time of the second quench in Fig.\ \ref{TwoStepWs0}. Parameters: $d=0.15$, $L=128$, $w_s=0$, and $T=0.1$. (Composition before initial equilibration at $T=0.7$: $p^A=0.470$ and $p^B=0.451$ for the slab liquid domain and $p^A=0.0247$ and $p^B=0.0955$ for the vapour. The slab extremes are placed at $18\%$ and $82\%$ of the horizontal simulation box side.) From top left to bottom right, the snapshots are taken at $t=0$, $2.32$, $7.10$, $17.61$, $22.65$, and $60.06$, counted from the moment of the quench to the final temperature.}
	\label{GridSSWs0SlabEqlbted}
\end{figure}

Fig.\ \ref{GridSSWs0SlabEqlbted} shows a deep quench simulation that replicates the physics seen above for $w_s=0$. We start from a slab-geometry liquid domain surrounded by a vapour phase,  \textit{equilibrated} at the first-quench temperature used for Fig.\ \ref{TwoStepWs0}. The initial liquid and vapour compositions as well as the other parameters therefore match those just before the second quench in Fig.\ \ref{TwoStepWs0}.
We then start the phase separation experiment by quenching to $T=0.1$, i.e.\ the second-quench temperature in Fig.\ \ref{TwoStepWs0}. We can clearly see that spinodal waves propagate from the slab-vapour interfaces towards the centre of the liquid. In the terminology used in ref.\ \citenum{Plapp99}, this is a consequence of surface modes, and this phenomenon can be viewed as a case of surface-directed spinodal decomposition.\cite{Lamellar,TangMingSurfaceEffects} The decomposition fronts are spaced according to our calculated spinodal length. The spinodal wave fronts turn into a succession of liquid and 
gas stripes; the latter eventually start to break into bubbles as dictated by the bulk modes. 

\begin{figure}[h]
	\centering
	\includegraphics[width=\columnwidth]{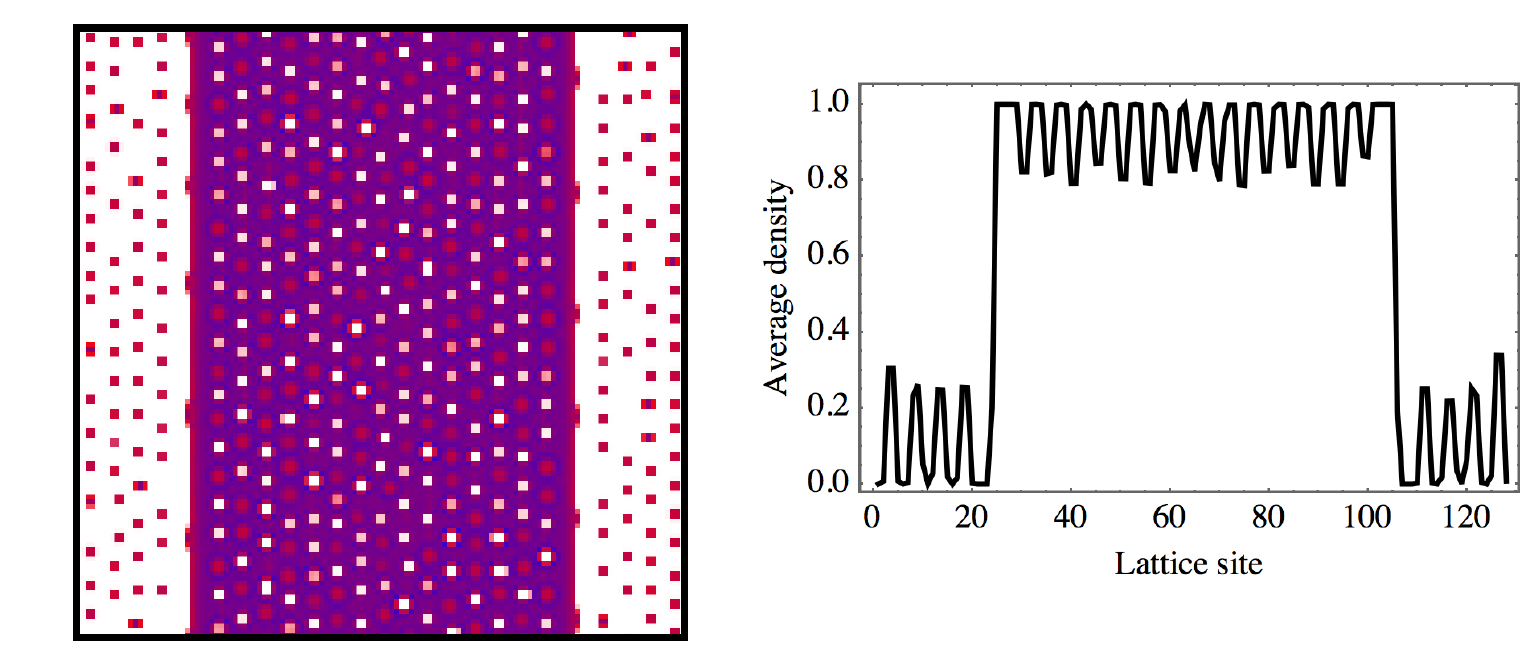}
	\caption{Left: as Fig.\ \ref{GridSSWs0SlabEqlbted} but at $t=86000$. Right: Corresponding average total density profile along cross-interface direction: for each horizontal point we plot the total density obtained by averaging over the vertical direction.}
	\label{SnapPlusProfileWs0SlabEqlbted}
\end{figure}

Note that the set of bubbles creates an ordered structure, i.e.\ the domains are fairly regularly spaced. The behaviour is similar to that obtained for the two-step quench simulations with polymer blends in ref.\ \citenum{ClarkeTwoStep}. Here, however, this regularity is also long-lived. This can be seen from Fig.\ \ref{SnapPlusProfileWs0SlabEqlbted}. We show on the left a snapshot of the system at a late time. On the right we show the corresponding average density profile along the cross-interface direction (i.e.\ normal to the slab-vapour interface); this will be useful for later comparison. Notice how spatially ordered the set of domains (plus long-lived patches) remains: even though the liquid is highly inhomogeneous in composition, the different structures retain their position for a long time as inherited from the original spinodal decomposition fronts.

\begin{figure}[h]
	\centering
	\includegraphics[width=\columnwidth]{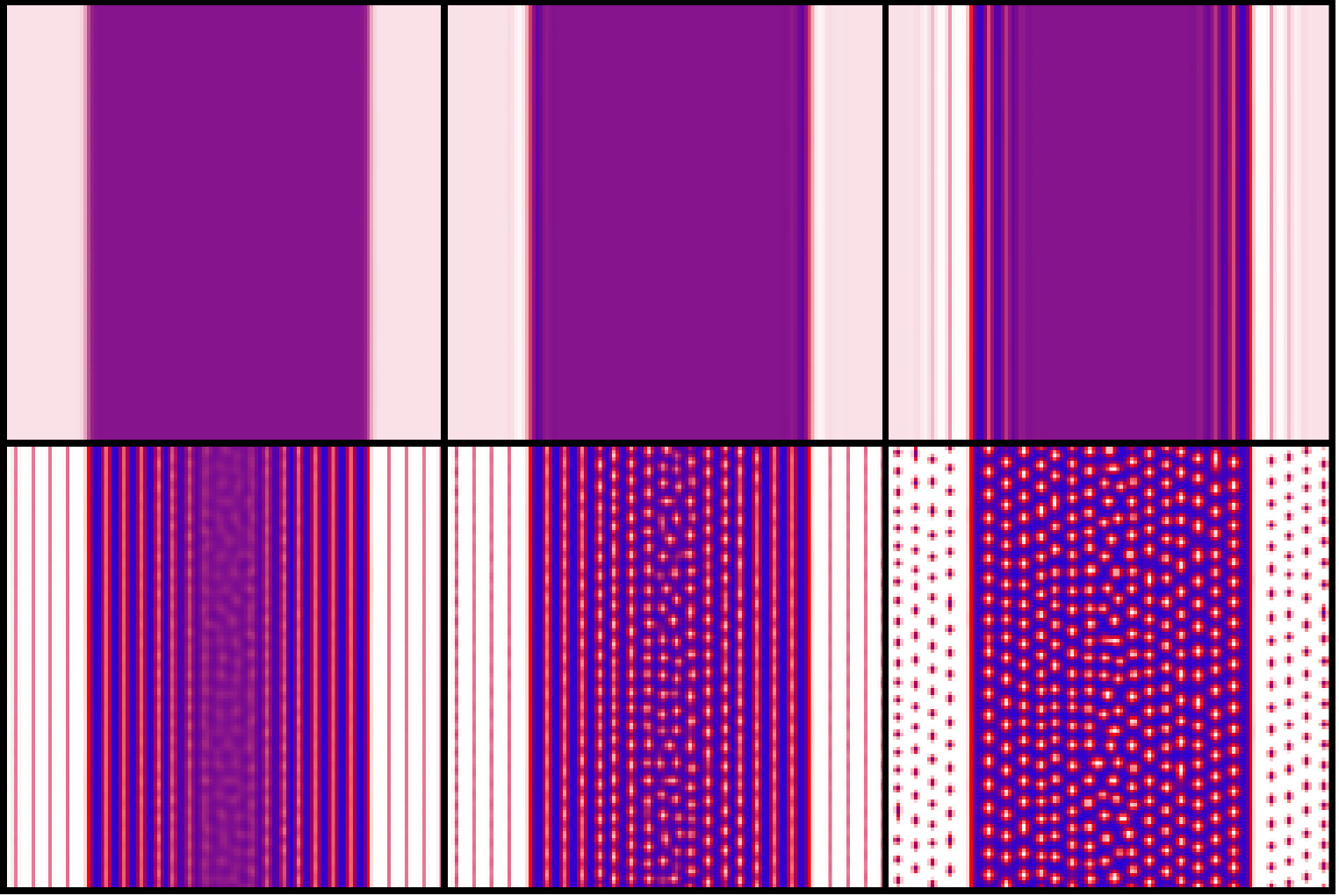}
	\caption{Analogue of Fig.\ \ref{GridSSWs0SlabEqlbted} for $w_s=1$. From top left to bottom right, the snapshots are taken at $t=0$, $1.89$, $6.41$, $14.93$, $18.49$, and $43.67$, counted from the moment of the quench to the final temperature.}
	\label{GridSSWs1SlabEqlbted}
\end{figure}

\begin{figure}[h]
	\centering
	\includegraphics[width=\columnwidth]{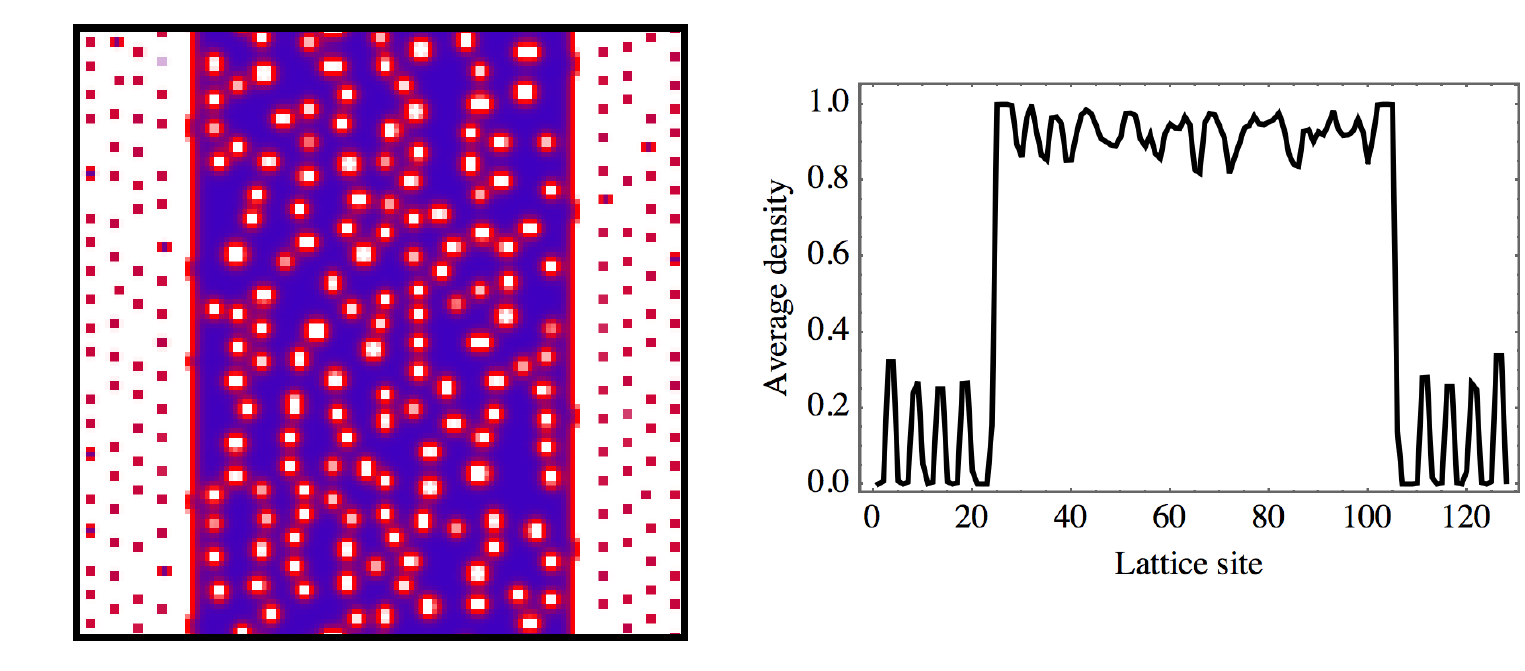}
	\caption{As Fig.\ \ref{SnapPlusProfileWs0SlabEqlbted} but for $w_s=1$. Data taken at $t=86000$.}
	\label{SnapPlusProfileWs1SlabEqlbted}
\end{figure}

The initial behaviour for $w_s=1$ shown in Fig.\ \ref{GridSSWs1SlabEqlbted} is structurally similar to that for $w_s=0$ except for much stronger fractionation in the homogeneous liquid (behind the decomposition front) and the interfaces. However, once the smallest bubbles start to collapse as we have seen in Section \ref{DeepQuench}, the regularity is destroyed gradually. This can been seen from the late-time snapshot and its corresponding average density profile in Fig.\ \ref{SnapPlusProfileWs1SlabEqlbted}.

\begin{figure}[h]
	\centering
	\includegraphics[width=\columnwidth]{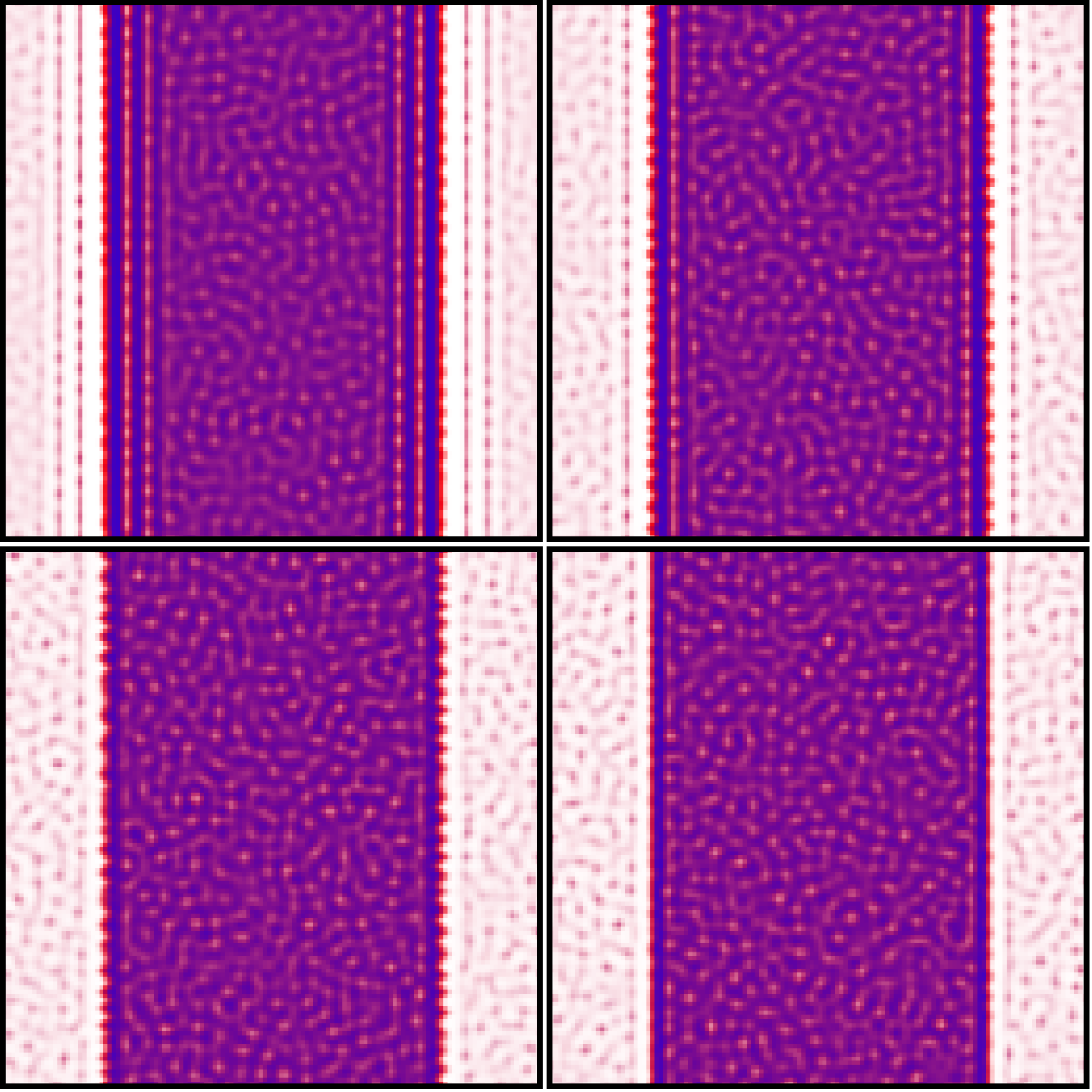}
	\caption{Parameters as Fig.\ \ref{GridSSWs1SlabEqlbted} but each early-time snapshot here corresponds to a run with a different strength of the noise added to the initial state, i.e.\ at the time of the final temperature quench. From top left to bottom right, the zero-mean initial noise has standard deviations $0.01\%$, $0.1\%$, $1\%$, and $1\%$ again, but in the bottom-right snapshot the slab-vapour interface has not been previously equilibrated. All snapshots are taken at the time within the spinodal regime where domains start to form (
$t=7.52$, $5.53$, $2.93$, and $3.32$ from top left to bottom right).
}
	\label{ThreeNoisesPlusNonEquil}
\end{figure}

The spatial regularity in the set of newly-formed domains is a direct consequence of the fact that phase separation is triggered via spinodal decomposition fronts propagating from the slab-vapour interfaces. The spinodal waves, in turn, occur because the initial higher-$T$ equilibration of the slab-vapour interfaces has removed all added fluctuations in the bulk phases---similarly to the double-quench simulations of a hydrodynamical model for polymeric blends in ref.\ \citenum{WangLi2011ViscousTwoStep}---leaving the slab-vapour interfaces as the sole source of fluctuations. That is, even though noise has been added to the slab-geometry configuration, this has been done prior to the initial equilibration, resulting in `noiseless' equilibrated phases, from which state no spinodal decomposition can occur except if started at the interface. In order to investigate the effects produced by this we have added noise back in at the moment of the final quench, as shown in Fig.\ \ref{ThreeNoisesPlusNonEquil}. Notice how the distinct noise strengths result in the spinodal waves being able to travel different distances before bulk spinodal modes reach their nonlinear regime. This is similar to what can be seen in the off-critical `A-B' vacancy-mediated phase separation simulations of binary alloys in ref.\ \citenum{Plapp99}. (As in that study we also performed simulations starting from a `droplet' geometry, which led to similar results.) It is worth pointing out that, because in the set-up of ref.~\citenum{Plapp99} the interaction energies are such that $\epsilon_{AA}=\epsilon_{BB}$, these `spinodal waves' can occur only in their off-critical case, where the overall system becomes richer in one of the particle species; this leads to the species chemical potentials being different, $\mu_A\neq \mu_B$, and consequently to a quick formation of interfaces that are richer in one component. In our case this situation arises by default as $\epsilon_{AA}\neq\epsilon_{BB}$.

The bottom-right snapshot in Fig.\ \ref{ThreeNoisesPlusNonEquil} is for a simulation with a \textit{non-equilibrated} initial slab configuration but with noise added as usual. Notice that, even though we do not start with previously formed smooth slab-vapour interfaces, the system (at the point where domains start to form) looks essentially the same as its equilibrated counterpart (bottom-left snapshot). This is because at this noise strength the interface (having a smooth density profile or a sharp density drop) is not the main source of fluctuations; instead, fluctuations start to grow everywhere simultaneously. (Careful inspection reveals one difference between the two plots on the bottom of Fig.\ \ref{ThreeNoisesPlusNonEquil}: in the case on the left the slab interface shows small undulations. These are generated because the initially equilibrated interface structure, with its densities intermediate between gas and liquid, becomes itself spinodally unstable after the second quench.) We point out that in ref.\ \citenum{Plapp99} it has been analytically derived that the penetration depth of the surface-directed structures depends logarithmically on the strength of the initial noise; this is consistent with what we see in Fig.\ \ref{ThreeNoisesPlusNonEquil}.

The above discussion shows the importance of considering noise in double quench theories, as otherwise the primary phases have no noise and therefore the primary domains would act as the sole sources of fluctuations. The exact noise strength will change with time, depending on temperature and composition, and this will dictate whether spinodal waves will develop or not, and consequently the structure of the set of secondary domains. As we have seen above, whether this structure is then long-lived will depend on the presence of slow-fractionation effects.

\subsection{Higher temperature}

We now turn our attention to secondary phase separation at higher temperatures rather than after deep quenches. Here we will choose a second quench temperature high enough for the primary gas phase not to undergo further phase separation while still  allowing the dense primary liquid to phase separate. In order to be able to increase the final temperature while keeping the liquid well within the spinodal region we will increase the polydispersity parameter $d$. Because Fig.\ \ref{ThreeNoisesPlusNonEquil} shows that by increasing the initial noise strength at the moment of the final quench one suppresses the surface-directed spinodal waves, we will in this higher temperature context simplify the simulations by ignoring effects from a previously formed smooth interface. This will be done by considering non-equilibrated, sharp slab-vapour interfaces for the initial state. The two situations should generally be equivalent if the bulk noise strength is high enough. With these choices we will be able to show the relevant physics in its simplest setting.

\begin{figure}[h]
	\centering
	\includegraphics[width=\columnwidth]{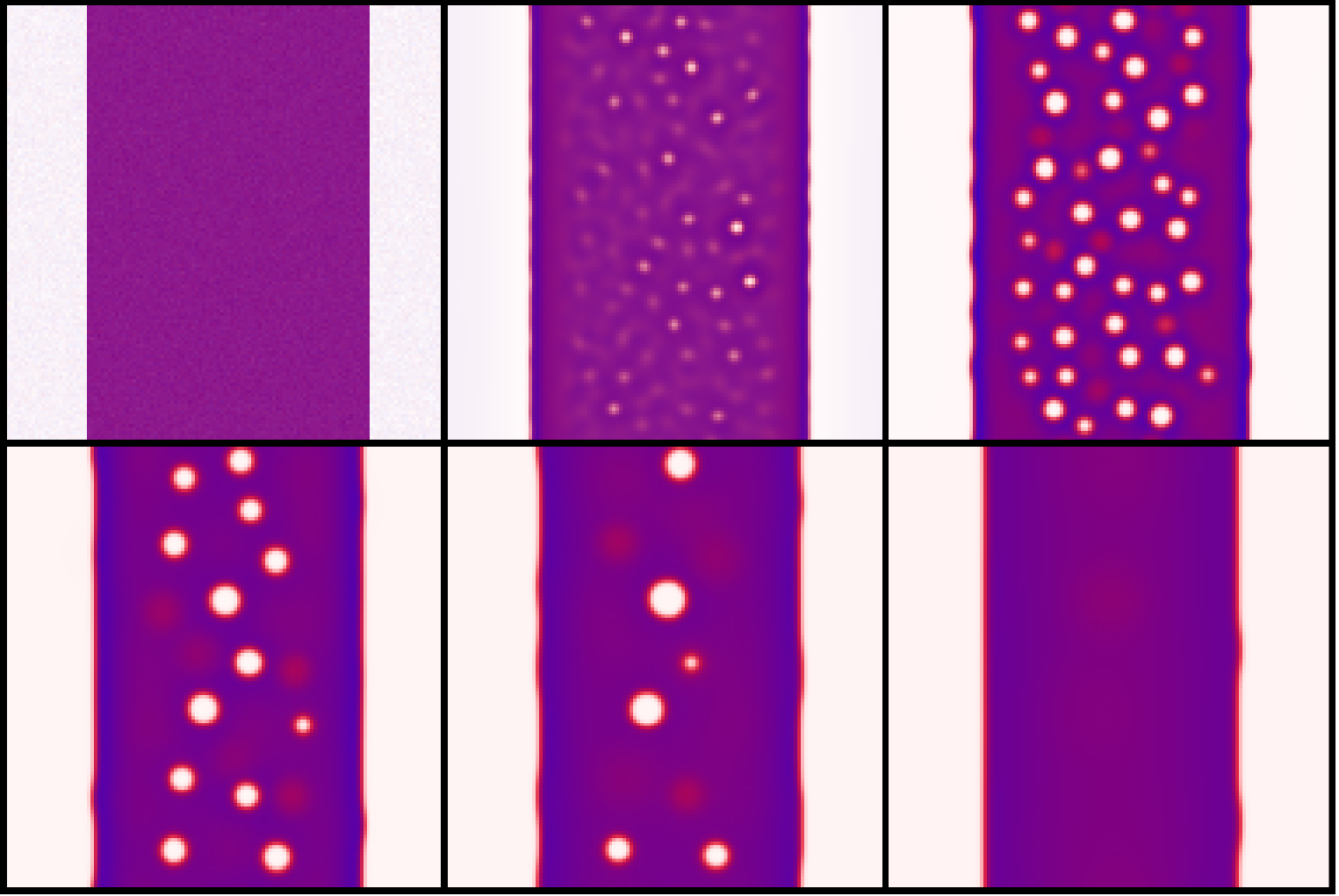}
	\caption{Time snapshots showing the local compositions throughout the system. We start from a slab-geometry liquid domain surrounded by vapour. Parameters: $d=0.25$, $L=128$, $w_s=0$, and $T=0.4$. Composition before quench: $p^A=p^B=0.45$ for the slab liquid domain and $p^A=p^B=0.03$ for the vapour. The slab extremes are initially placed at $18\%$ and $82\%$ of the horizontal  simulation box side. From top left to bottom right, the snapshots are taken at $t=0$, $131$, $2089$, $18445$, $41111$, and $122065$.
}
	\label{GridSSWs0SlabHigherT}
\end{figure}

Fig.\ \ref{GridSSWs0SlabHigherT} shows a simulation for $w_s=0$ where both the initial phases, the dense liquid and the dilute vapour, contain equal amounts of $A$ and $B$ particles. After the temperature quench a smooth interface with the outside vapour is formed quickly and after some time spinodal decomposition can be seen to start in multiple locations simultaneously. In parallel, a layer rich in $A$'s is formed rapidly on the liquid side of the interface.
The formation of this $A$-rich zone has two contributing causes: (i) generally in our system interfaces contain mainly $B$ particles so the formation of the interface will transiently deplete $B$'s from the liquid outside the interface;
(ii) if the new gas phase is significantly enriched in $B$'s, or conversely depleted in $A$'s compared to the initial phase, then the surplus $A$ particles will diffuse into the liquid phase and create an excess of $A$'s there. 

Looking at the longer-time regime in Fig.\ \ref{GridSSWs0SlabHigherT} one notices first that overall there is a density increase in the liquid, making the slab thinner as time passes. As previously we can also see $B$-rich long-lived patches as remnants of the interfaces of `dead' bubbles, but here the $A$-rich layer constitutes a further long-lived structure. Over time it gradually becomes thicker but also less fractionated (and so less distinguishable from the bulk liquid); ultimately it dissolves and so do the bubbles, with the liquid then close to a homogeneous state.

To understand the differences to deep quenches as in Section \ref{revisited}, where no equivalent of the $A$-rich layer was observed, we note that the mechanism (ii) described above cannot contribute to the formation of such a layer in deep quenches: there are then essentially no particles remaining in the gas phase as even at equilibrium the gas will be extremely dilute. Also, the diffusive exchange with the surrounding gas has to have time to form an $A$-rich zone before spinodal fluctuations grow; this situation arises after performing a shallow quench, in which case the spinodal dynamics is generally slower---for both $w_s=0$ and $w_s=1$ cases---than it is for a deep quench. (This relation between the spinodal timescales remains even when taking into account the different value of $d$ here, as we have checked by calculating the spinodal rates explicitly.)

\begin{figure}[h]
	\centering
	\includegraphics[width=\columnwidth]{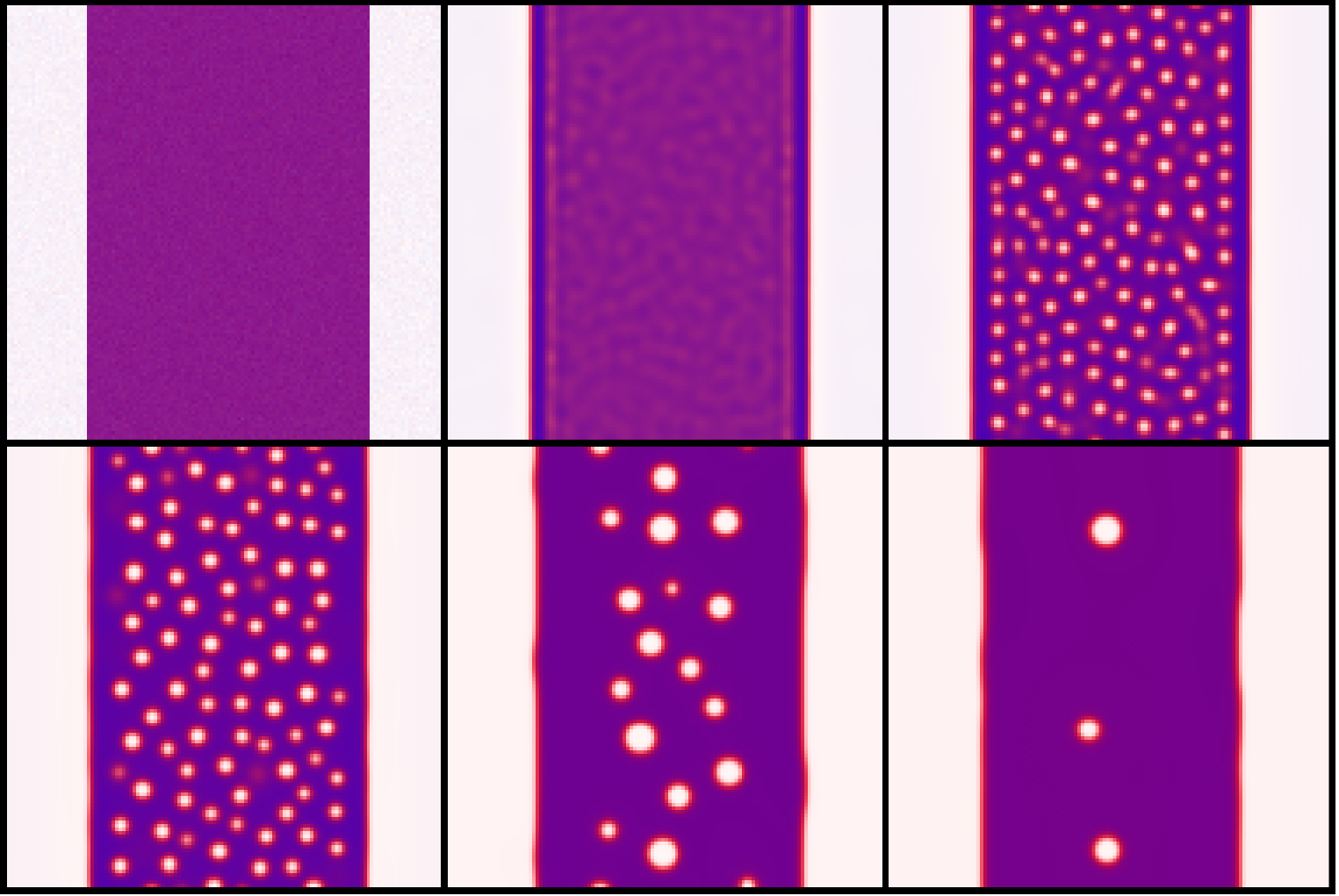}
	\caption{As Fig.\ \ref{GridSSWs0SlabHigherT} but for $w_s=1$. From top left to bottom right, the snapshots are taken at $t=0$, $116$, $1703$, $15791$, $36479$, and $87289$.
}
	\label{GridSSWs1SlabHigherT}
\end{figure}

For a shallow quench with $w_s=1$ shown in Fig.\ \ref{GridSSWs1SlabHigherT} we can initially see an $A$-rich layer similar to the case $w_s=0$. Its thickness is related to the spinodal length, which leads to domains being able to grow only above a certain minimum distance from the slab-vapour interface. (Notice that fluctuations start to grow simultaneously everywhere within a region of composition nearly $50$--$50\%$.) In contrast to the $w_s=0$ case, however, the liquid quickly becomes homogeneous. Since there is initially a significant concentration of bubble interfaces---which are enriched in $B$ particles---the bulk liquid ends up being homogeneously slightly enriched in $A$'s for a long period of time, as close inspection of Fig.\ \ref{GridSSWs1SlabHigherT} reveals. Eventually the bulk liquid reaches a less fractionated composition and the bubbles are reabsorbed.

\begin{figure}[h]
	\centering
	\includegraphics[width=\columnwidth]{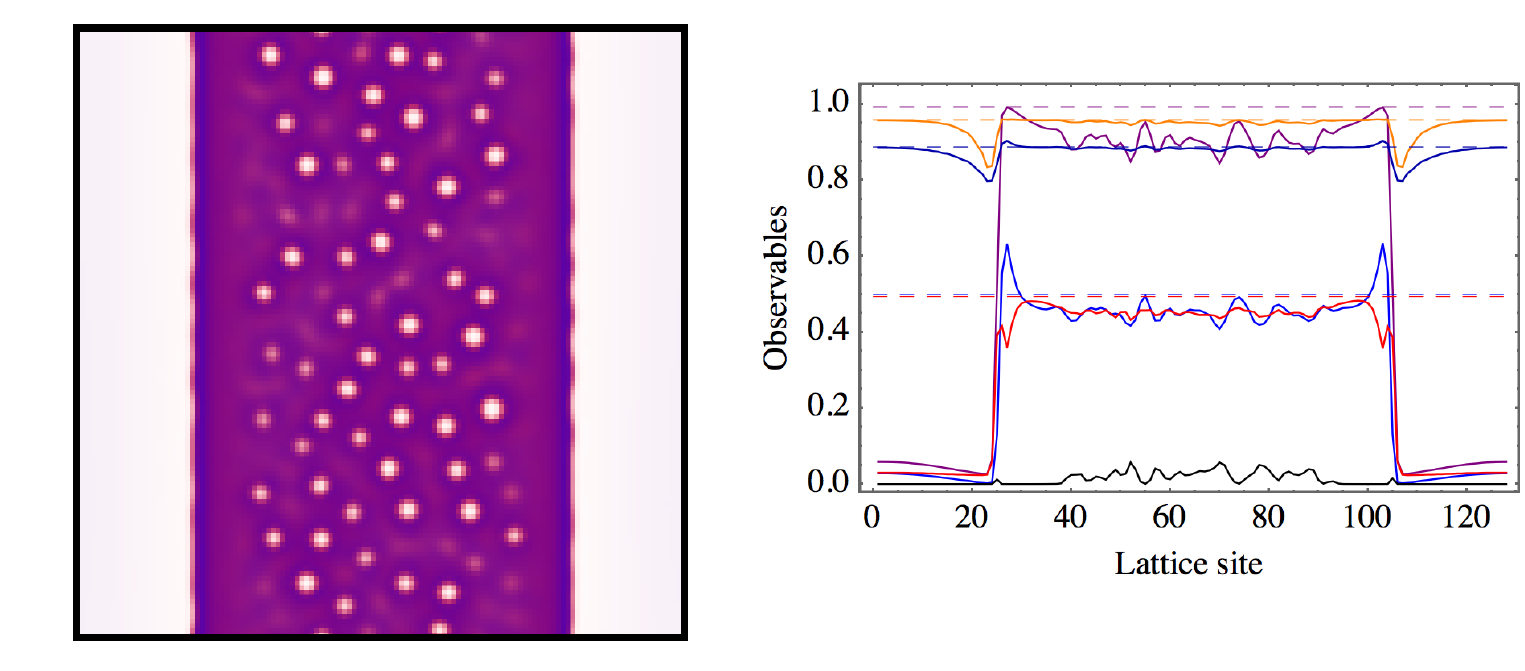}
	\caption{Left: as Fig.\ \ref{GridSSWs0SlabHigherT} but at $t=172$. Right: Profiles of several observables along the cross-interface direction (same time). All profiles are obtained by averaging over the vertical direction at the given horizontal position. Observables (solid lines)---also listed in Table \ref{TableObservables}---from top to bottom along the $y$-axis: composition-dependent average annealed spinodal density (orange), composition-dependent average quenched spinodal density (darker blue), average total density (purple), average $p^B$ (red), average $p^A$ (lighter blue), and variance of total density (black). Dashed lines, from top to bottom along the $y$-axis: liquid equilibrium total density (purple), overall liquid annealed spinodal density (orange), overall liquid quenched spinodal density (darker blue), liquid equilibrium $p^A$ (lighter blue), and liquid equilibrium $p^B$ (red). All spinodal densities are from the high-density branch.}
	\label{DeadZone}
\end{figure}

\begin{figure}[h]
	\centering
	\includegraphics[width=\columnwidth]{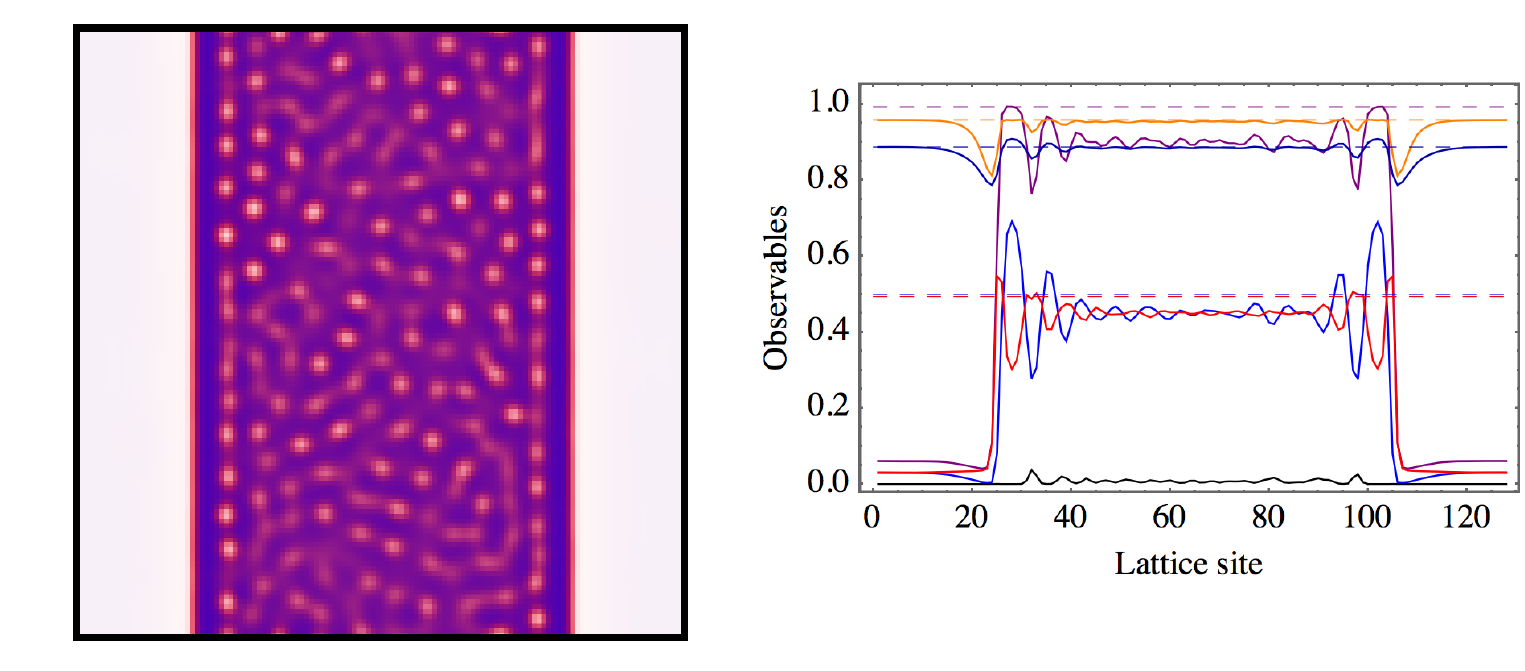}
	\caption{As Fig.\ \ref{DeadZone} but for $w_s=1$ and at $t=50$. Observables plotted on the right-hand side are also as in Fig \ref{DeadZone}; see Table \ref{TableObservables}.}
	\label{NoDeadZone}
\end{figure}

We notice further an interesting phenomenon that is only possible at high enough temperatures. In the case $w_s=0$, for times where the bubbles are still in their initial growth regime (see Fig.\ \ref{DeadZone}), we can observe the existence of a `dead zone', i.e.\ a stripe where no composition fluctuations grow, located next to each $A$-rich layer. As the spinodal dynamics is quite slow for $w_s=0$, we see that by the time domains start to form, a wider region of liquid has established itself near the interface; it mostly has density close to the equilibrium value and is thus above the density of even the (upper) annealed spinodal curve. This region is therefore no longer unstable to small composition fluctuations, so these will be damped rather than grow into bubbles. The phenomenon can be seen also from the two humps in the average total density curve near the slab boundaries on the right-hand side of Fig.\ \ref{DeadZone}; they show that the region has become sufficiently dense---and if we move towards the centre of the system from these humps we still remain (over some length) at total densities that are very close to the annealed spinodal, therefore making the growth rates very small and leading effectively to a wide `dead zone'. This can also be seen from the variance of the total density: it is zero in a wide region near the boundary. This is another interesting manifestation of Warren's scenario: the spinodal dynamics is slow because we are mostly outside  at least the quenched spinodal (see right-hand side of Fig.\ \ref{DeadZone} again, where we show the composition-dependent spinodal densities as cross-interface profiles). Therefore in the liquid region near the interface the fast process of density equilibration wins and stabilizes the system before fluctuations can grow. In a one-component system this cannot happen as the timescale of the spinodal dynamics is directly set by the density equilibration. As expected, the effect is destroyed by switching on particle swaps ($w_s=1$, see Fig.\ \ref{NoDeadZone}) as this significantly accelerates the spinodal dynamics.

\begin{table}[]
	\centering
	\vspace{0.5cm}
	\begin{tabular}{|l|l|}
		\hline
		\textbf{Observables (solid lines) }                                                      & \textbf{Colour}       \\ \hline
		Average annealed spinodal density & Orange       \\ \hline
		Average quenched spinodal density              & Darker blue  \\ \hline
		Average total density                                                           & Purple       \\ \hline
		Average $p^B$                                                                   &  Red \\ \hline
		Average $p^A$                                                                   & Lighter blue  \\ \hline
		Variance of total density                                                       & Black        \\ \hline
	\end{tabular}
	\caption{Observables (solid lines) plotted in Fig.\ \ref{DeadZone} (right), listed in order from top to bottom along the $y$-axis.}
	\label{TableObservables}
\end{table}

\section{Secondary quench into three-phase region}
\label{threephase}

We return finally to the double quench case in bulk geometry, with the second quench now chosen deep enough to enter a three-phase region. While the kinetics of phase separation after quenches into a three-phase coexistence region has been studied previously,\cite{tafa2001kinetics} the range of parameters investigated was relatively limited and to the best of our knowledge the double quench case has not been considered in the literature. Here we consider a higher $d$ than in the discussion for the secondary quench into the two-phase region; therefore we will be able to choose higher final temperatures while still reaching the three-phase region.

\begin{figure}[h]
	\centering
	\includegraphics[width=\columnwidth]{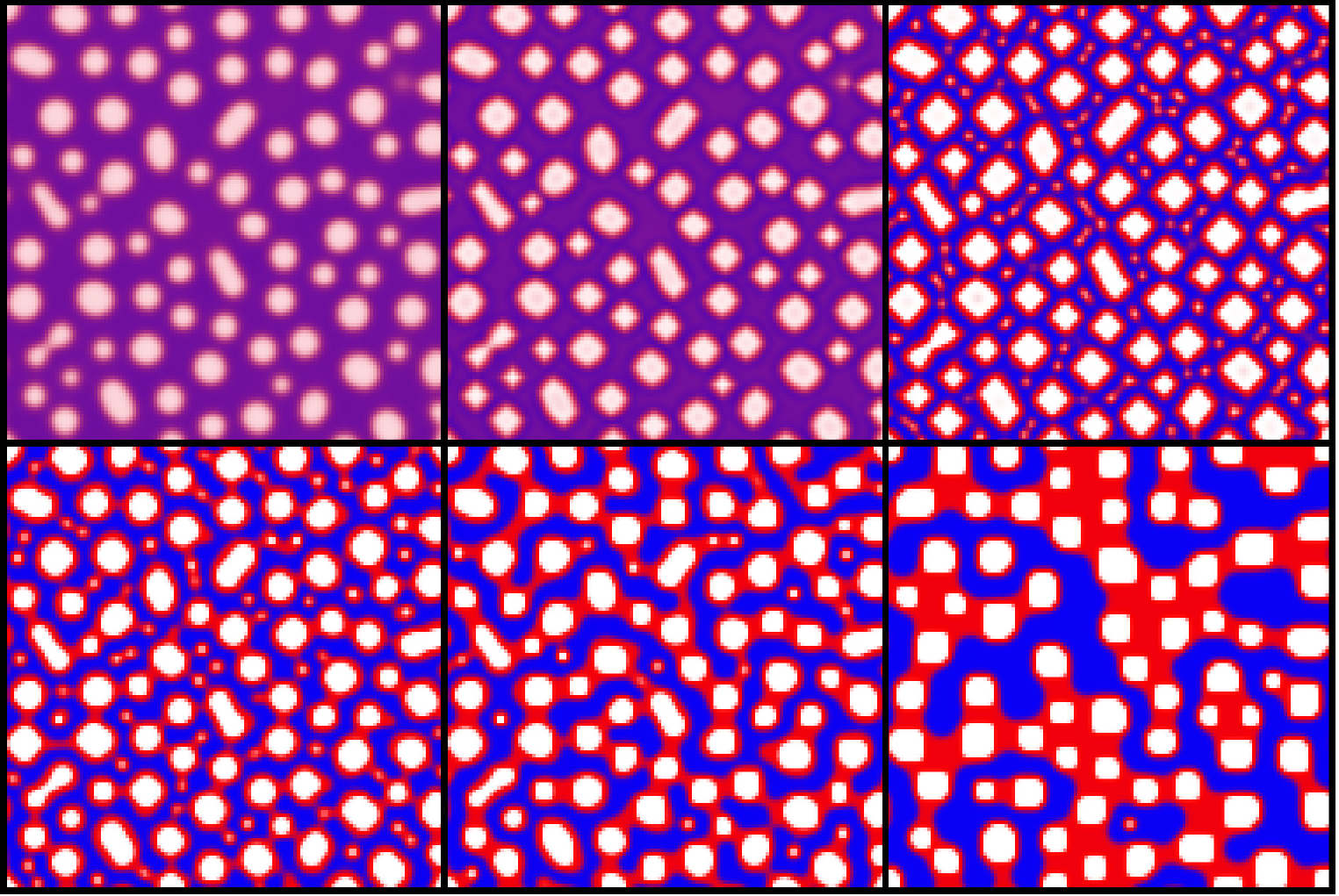}
	\caption{As Fig.\ \ref{TwoStepWs1} but for a second quench into the three-phase coexistence region. Parameters: $p^A=p^B=0.375$, $T_{1}=0.7$, $T_2=0.15$, $d=0.25$, $w_s=1$, and $L=128$. From top left to bottom right, the snapshots are taken at $t=1500$, $1501$, $1510$, $1597$, $1914$, and $225702$. The second quench was performed at $t=1500$.}
	\label{ThreePhaseFigureWs1}
\end{figure}

Fig.\ \ref{ThreePhaseFigureWs1} shows our results for a double quench experiment into the three-phase region. In this case the fast dynamics ($w_s=1$) induces an interesting wetting phenomenon whereby the primary domains eventually become connected via strongly fractionated interfaces that have been formed from the secondary bubbles. This is different from the wetting `bridges' described in Section \ref{DeepQuench} (which are a kinetic effect). Indeed, here these `filaments' persist (and thicken) due to a thermodynamic driving force: they eventually turn into a third, $B$-rich phase. 

\begin{figure}[h]
	\centering
	\includegraphics[width=\columnwidth]{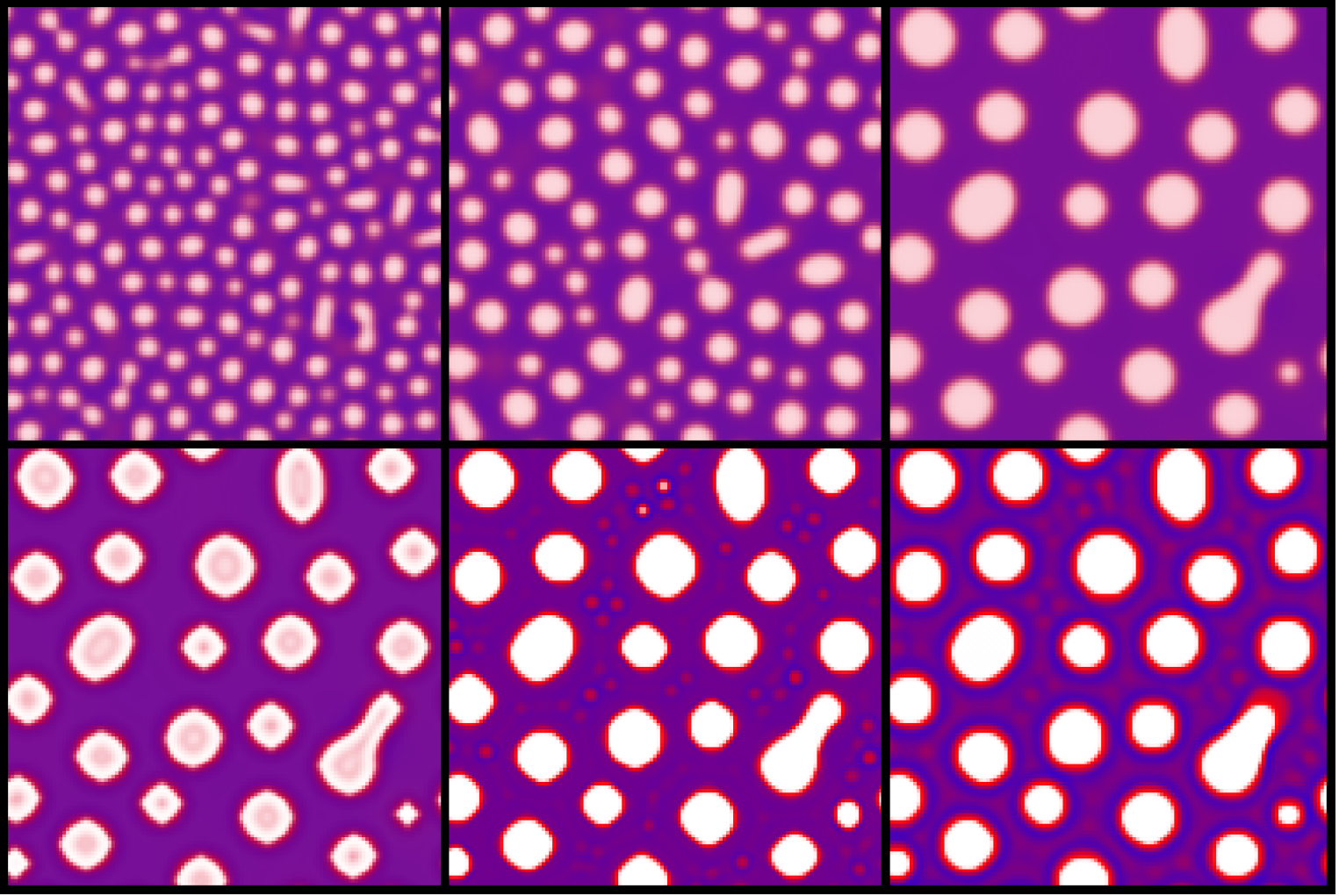}
	\caption{Second quench into a higher temperature in the three-phase region, for $w_s=0$. Other parameters are: $p^A=p^B=0.375$, $T_{1}=0.7$, $T_2=0.20$, $d=0.25$, and $L=128$. From top left to bottom right, the snapshots are taken at $t=349$, $1992$, $20000$, $20004$, $26177$, and $408307$. The second quench was performed at $t=20000$.}
	\label{ThreePhaseFigureWs0HigherT}
\end{figure}

\begin{figure}[h]
	\centering
	\includegraphics[width=\columnwidth]{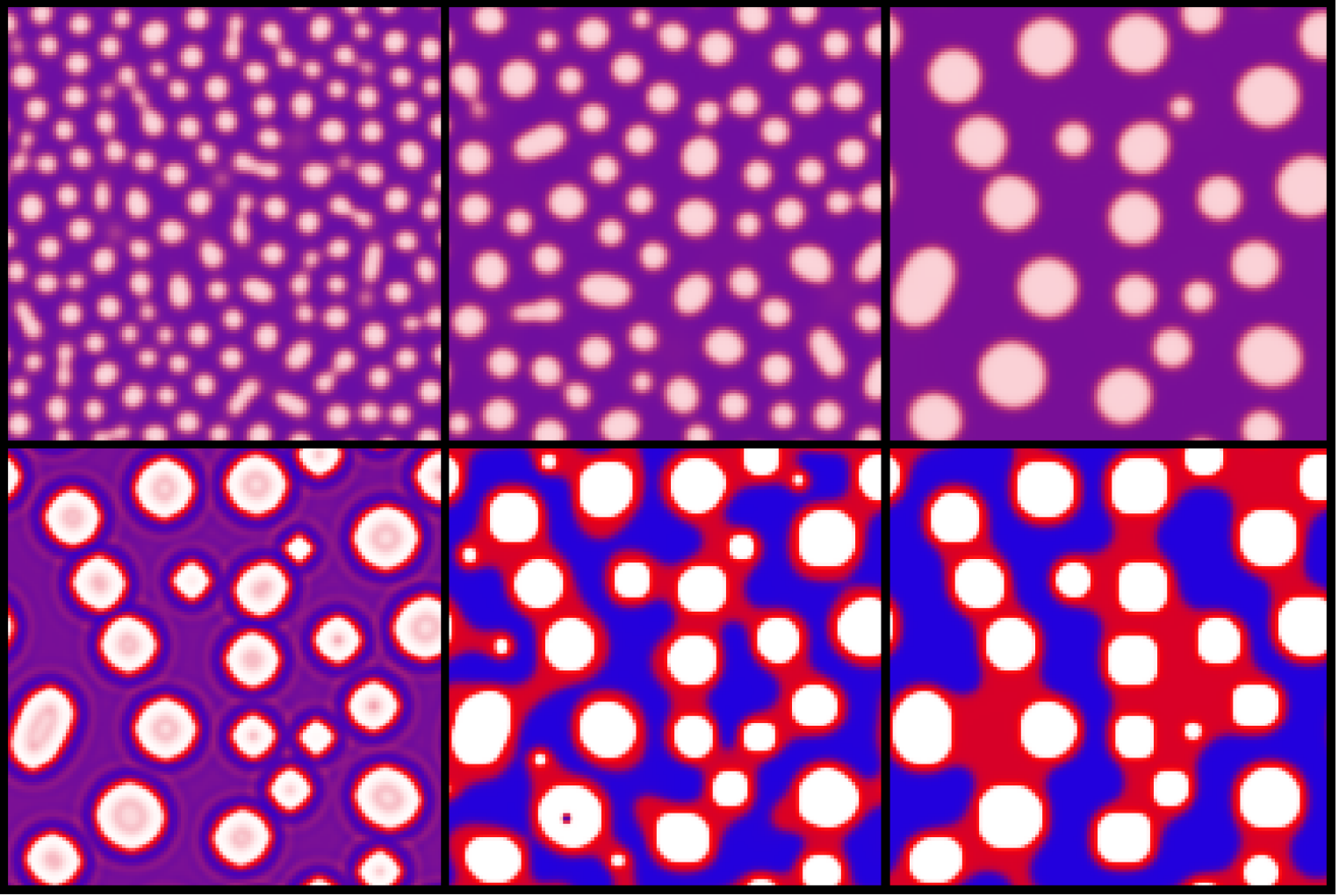}
	\caption{As Fig. \ref{ThreePhaseFigureWs0HigherT}, but for $w_s=1$. Parameters: $p^A=p^B=0.375$, $T_{1}=0.7$, $T_2=0.20$, $d=0.25$, $w_s=1$, and $L=128$. From top left to bottom right, the snapshots are taken at $t=326$, $1816$, $20001$, $20005$, $24515$, and $190388$. The secondary quench was performed at $t=20000$.}
	\label{ThreePhaseFigureWs1HigherT}
\end{figure}

Figs.\ \ref{ThreePhaseFigureWs0HigherT} and \ref{ThreePhaseFigureWs1HigherT} show simulations for the cases $w_s=0$ and $w_s=1$, respectively, performed for much larger second quench time $t_2$. The second quench temperature $T_2$ is also significantly higher $T_2$, but still low enough to take the system into its three-phase region. The results in the figures indicate that, until the third-phase filaments start to form and connect primary bubbles, the morphology evolves in a way that is generally similar to the two-phase cases in Section \ref{DeepQuench} (see Figs. \ref{TwoStepWs0} and \ref{TwoStepWs1}). Specifically, for $w_s=0$ the behaviour for the two-phase and three-phase cases should be qualitatively the same until the liquid phase becomes homogeneous enough to form filaments of the $B$-rich phase: in the final snapshot in Fig.\ \ref{ThreePhaseFigureWs0HigherT} we can already see a clear broadening of the interfaces as the third phase starts to be formed. We can even see that one of the primary bubbles, which has a protuberance in addition to its round shape (from the merger between two rounded domains), begins to form a broader interface, possibly an incipient filament, more quickly than the other bubbles. This could be interesting to explore in future work, by simulating a three-phase final quench starting from a controlled geometry of primary domains. At any rate, in this $w_s=0$ case the formation of filaments will generally take much longer; therefore it would be interesting to study the final morphology at even longer times in future studies.

\section{Conclusions}
\label{ConclusionsDouble}
In this work we have shown how composition heterogeneities driven by kinetic effects arise in a variety of forms in phase-separating colloidal mixtures. Our results supplement ref.~\citenum{PabloPeter1} and constitute additional theoretical evidence that polydisperse phase separation can proceed rather differently from its monodisperse counterpart due to crowding effects on the kinetics of compositions across the system. Here we have looked at the phase separation process following a second temperature quench into the two- and three-phase regions. Secondary domains grow inside the primary phases, but they never get to coarsen without limit: they are eventually reabsorbed through the primary domain interfaces, as in the incompressible polymeric case of ref.\ \citenum{ClarkeTwoStep}.

Here, however, for a deep second quench, the coarsening of the primary domains is interrupted, and the  morphology of the liquid formed by secondary phase separation changes with the kinetics. Also, we investigated the case of a slab surrounded by vapour, which can be interpreted as equivalent to waiting an infinite amount of time before performing the second quench into the two-phase region in a two-step quench experiment. Spinodal waves appear as a surface-directed phenomenon, and the decomposition fronts eventually break into secondary bubbles. In the case with slow composition changes, this leads to \textit{long-lived} regularity in the spatial arrangement of the secondary domains. As already pointed out in ref.\ \citenum{ClarkeTwoStep}, this can inspire strategies to develop regular morphologies that may have unusual physical properties. The fact that in our case this regularity is long-lived generates further opportunities to obtain new structures in soft matter systems. The formation of spinodal waves from the initially equilibrated slab-vapour interfaces is affected by the amount of noise existent at the time of the final quench. 

At higher temperatures we have seen that a layer rich in the strongest-interacting species can form because the diffusive exchange with the surrounding gas has time to form such a zone before spinodal fluctuations grow. This is a result of the depletion of the most weakly interacting species, which flow into the gas. For $w_s=0$ this layer is a significantly more long-lived composition heterogeneity. In this case we also showed the existence of a `dead zone' where spinodal decomposition does not occur. This is a further manifestation of Warren's two-stage scenario.

Finally we presented our simulations for the case of a second quench into the three-phase region, where we observed the formation of filaments out of the primary domains, mediated by the secondary domains; these filaments wet and connect the primary domains and eventually turn into the third phase.

In future work on deep secondary quenches we would like to investigate how the time length of the coarsening-interruption plateau depends on effects of slow composition dynamics, and how the coarsening is restored as we approach the LS law. For the higher-$T$ slab case, one could check how the dead zone is affected by nucleation and growth. In this case, even though spinodal decomposition with respect to species `sizes' (i.e.\ composition equilibration via spinodal decomposition) is not present in that region there could still be a situation where nucleation and growth takes over the phase separation process, effectively destroying the dead zone. Studying nucleation and growth dynamics in this way would require including noise throughout the dynamics, rather than just in the initial condition for the secondary quench; extending the model in this way would then allow one also to assess what effects noise has on the standard spinodal dynamics. In the three-phase double quench case, we anticipate further interesting phenomena in the evolution of the morphology for $w_s=0$, on timescales longer than we have been able to access here. Also, extending the three-phase dynamical problem to controlled geometries of the primary interfaces that include e.g.\ protuberances could constitute an effective method for analysing in a simple way the dynamics of filament formation, at least for $w_s=1$. More specifically, it would be interesting to look into the dynamical process by which two primary liquid domains become connected. Finally, it will also be worth seeing how the effects investigated here manifest themselves in off-lattice models.

\section*{Conflicts of interest}
There are no conflicts to declare.

\section*{Acknowledgements}
PdC acknowledges financial support from CNPq, Conselho Nacional de Desenvolvimento Cient\'{i}fico e Tecnol\'{o}gico -- Brazil (GDE 202399/2014-1). PS acknowledges the stimulating research environment provided by the EPSRC Centre for Doctoral Training in Cross-Disciplinary Approaches to Non-Equilibrium Systems (CANES, EP/L015854/1).



\balance

\renewcommand\refname{Notes and references}

\bibliography{Polydispersity.bib} 

\providecommand*{\mcitethebibliography}{\thebibliography}
\csname @ifundefined\endcsname{endmcitethebibliography}
{\let\endmcitethebibliography\endthebibliography}{}
\begin{mcitethebibliography}{40}
\providecommand*{\natexlab}[1]{#1}
\providecommand*{\mciteSetBstSublistMode}[1]{}
\providecommand*{\mciteSetBstMaxWidthForm}[2]{}
\providecommand*{\mciteBstWouldAddEndPuncttrue}
  {\def\EndOfBibitem{\unskip.}}
\providecommand*{\mciteBstWouldAddEndPunctfalse}
  {\let\EndOfBibitem\relax}
\providecommand*{\mciteSetBstMidEndSepPunct}[3]{}
\providecommand*{\mciteSetBstSublistLabelBeginEnd}[3]{}
\providecommand*{\EndOfBibitem}{}
\mciteSetBstSublistMode{f}
\mciteSetBstMaxWidthForm{subitem}
{(\emph{\alph{mcitesubitemcount}})}
\mciteSetBstSublistLabelBeginEnd{\mcitemaxwidthsubitemform\space}
{\relax}{\relax}

\bibitem[Poon(2002)]{poon2002physics}
W.~Poon, \emph{Journal of Physics: Condensed Matter}, 2002, \textbf{14},
  R859\relax
\mciteBstWouldAddEndPuncttrue
\mciteSetBstMidEndSepPunct{\mcitedefaultmidpunct}
{\mcitedefaultendpunct}{\mcitedefaultseppunct}\relax
\EndOfBibitem
\bibitem[Van~den Pol \emph{et~al.}(2009)Van~den Pol, Petukhov, Thies-Weesie,
  Byelov, and Vroege]{van2009experimental}
E.~Van~den Pol, A.~Petukhov, D.~Thies-Weesie, D.~Byelov and G.~Vroege,
  \emph{Physical Review Letters}, 2009, \textbf{103}, 258301\relax
\mciteBstWouldAddEndPuncttrue
\mciteSetBstMidEndSepPunct{\mcitedefaultmidpunct}
{\mcitedefaultendpunct}{\mcitedefaultseppunct}\relax
\EndOfBibitem
\bibitem[Stuart \emph{et~al.}(1980)Stuart, Scheutjens, and
  Fleer]{stuart1980polydispersity}
M.~A.~C. Stuart, J.~M. H.~M. Scheutjens and G.~J. Fleer, \emph{Journal of
  Polymer Science: Polymer Physics Edition}, 1980, \textbf{18}, 559--573\relax
\mciteBstWouldAddEndPuncttrue
\mciteSetBstMidEndSepPunct{\mcitedefaultmidpunct}
{\mcitedefaultendpunct}{\mcitedefaultseppunct}\relax
\EndOfBibitem
\bibitem[Evans \emph{et~al.}(1998)Evans, Fairhurst, and
  Poon]{evans1998universal}
R.~Evans, D.~Fairhurst and W.~Poon, \emph{Physical Review Letters}, 1998,
  \textbf{81}, 1326\relax
\mciteBstWouldAddEndPuncttrue
\mciteSetBstMidEndSepPunct{\mcitedefaultmidpunct}
{\mcitedefaultendpunct}{\mcitedefaultseppunct}\relax
\EndOfBibitem
\bibitem[Wilding \emph{et~al.}(2008)Wilding, Sollich, and
  Buzzacchi]{PhysRevE.77.011501}
N.~B. Wilding, P.~Sollich and M.~Buzzacchi, \emph{Physical Review E}, 2008,
  \textbf{77}, 011501\relax
\mciteBstWouldAddEndPuncttrue
\mciteSetBstMidEndSepPunct{\mcitedefaultmidpunct}
{\mcitedefaultendpunct}{\mcitedefaultseppunct}\relax
\EndOfBibitem
\bibitem[Vi{\'e}ville \emph{et~al.}(2011)Vi{\'e}ville, Tanty, and
  Delsuc]{vieville2011polydispersity}
J.~Vi{\'e}ville, M.~Tanty and M.-A. Delsuc, \emph{Journal of Magnetic
  Resonance}, 2011, \textbf{212}, 169--173\relax
\mciteBstWouldAddEndPuncttrue
\mciteSetBstMidEndSepPunct{\mcitedefaultmidpunct}
{\mcitedefaultendpunct}{\mcitedefaultseppunct}\relax
\EndOfBibitem
\bibitem[Auer and Frenkel(2001)]{auer2001suppression}
S.~Auer and D.~Frenkel, \emph{Nature}, 2001, \textbf{413}, 711--713\relax
\mciteBstWouldAddEndPuncttrue
\mciteSetBstMidEndSepPunct{\mcitedefaultmidpunct}
{\mcitedefaultendpunct}{\mcitedefaultseppunct}\relax
\EndOfBibitem
\bibitem[Belli \emph{et~al.}(2011)Belli, Patti, Dijkstra, and
  Van~Roij]{belli2011polydispersity}
S.~Belli, A.~Patti, M.~Dijkstra and R.~Van~Roij, \emph{Physical Review
  Letters}, 2011, \textbf{107}, 148303\relax
\mciteBstWouldAddEndPuncttrue
\mciteSetBstMidEndSepPunct{\mcitedefaultmidpunct}
{\mcitedefaultendpunct}{\mcitedefaultseppunct}\relax
\EndOfBibitem
\bibitem[Sollich(2005)]{sollich2005nematic}
P.~Sollich, \emph{The Journal of Chemical Physics}, 2005, \textbf{122},
  214911\relax
\mciteBstWouldAddEndPuncttrue
\mciteSetBstMidEndSepPunct{\mcitedefaultmidpunct}
{\mcitedefaultendpunct}{\mcitedefaultseppunct}\relax
\EndOfBibitem
\bibitem[Rogosi\'c \emph{et~al.}(1996)Rogosi\'c, Mencer, and
  Gomzi]{ROGOSIC19961337}
M.~Rogosi\'c, H.~Mencer and Z.~Gomzi, \emph{European Polymer Journal}, 1996,
  \textbf{32}, 1337 -- 1344\relax
\mciteBstWouldAddEndPuncttrue
\mciteSetBstMidEndSepPunct{\mcitedefaultmidpunct}
{\mcitedefaultendpunct}{\mcitedefaultseppunct}\relax
\EndOfBibitem
\bibitem[Sollich(2002)]{PeterReview}
P.~Sollich, \emph{Journal of Physics: Condensed Matter}, 2002, \textbf{14},
  R79\relax
\mciteBstWouldAddEndPuncttrue
\mciteSetBstMidEndSepPunct{\mcitedefaultmidpunct}
{\mcitedefaultendpunct}{\mcitedefaultseppunct}\relax
\EndOfBibitem
\bibitem[de~Castro and Sollich(2018)]{PabloPeter2}
P.~de~Castro and P.~Sollich, \emph{The Journal of Chemical Physics}, 2018,
  \textbf{149}, 204902\relax
\mciteBstWouldAddEndPuncttrue
\mciteSetBstMidEndSepPunct{\mcitedefaultmidpunct}
{\mcitedefaultendpunct}{\mcitedefaultseppunct}\relax
\EndOfBibitem
\bibitem[de~Castro~Melo(2019)]{decastro2019}
P.~S. de~Castro~Melo, \emph{Phase separation of polydisperse fluids},
  2019\relax
\mciteBstWouldAddEndPuncttrue
\mciteSetBstMidEndSepPunct{\mcitedefaultmidpunct}
{\mcitedefaultendpunct}{\mcitedefaultseppunct}\relax
\EndOfBibitem
\bibitem[Zhang \emph{et~al.}(2019)Zhang, Peng, Zhang, Lei, Yao, and
  Wang]{zhang2019temperature}
L.~Zhang, Y.~Peng, L.~Zhang, X.~Lei, W.~Yao and N.~Wang, \emph{RSC Advances},
  2019, \textbf{9}, 10670--10678\relax
\mciteBstWouldAddEndPuncttrue
\mciteSetBstMidEndSepPunct{\mcitedefaultmidpunct}
{\mcitedefaultendpunct}{\mcitedefaultseppunct}\relax
\EndOfBibitem
\bibitem[de~Castro and Sollich(2017)]{PabloPeter1}
P.~de~Castro and P.~Sollich, \emph{Physical Chemistry Chemical Physics}, 2017,
  \textbf{19}, 22509--22527\relax
\mciteBstWouldAddEndPuncttrue
\mciteSetBstMidEndSepPunct{\mcitedefaultmidpunct}
{\mcitedefaultendpunct}{\mcitedefaultseppunct}\relax
\EndOfBibitem
\bibitem[B.~Warren(1999)]{A809828J}
P.~B.~Warren, \emph{Physical Chemistry Chemical Physics}, 1999, \textbf{1},
  2197--2202\relax
\mciteBstWouldAddEndPuncttrue
\mciteSetBstMidEndSepPunct{\mcitedefaultmidpunct}
{\mcitedefaultendpunct}{\mcitedefaultseppunct}\relax
\EndOfBibitem
\bibitem[McLeish \emph{et~al.}(2003)McLeish, Cates, Higgins, Olmsted, Cates,
  Vollmer, Wagner, and Vollmer]{CatesRamp}
T.~C.~B. McLeish, M.~E. Cates, J.~S. Higgins, P.~D. Olmsted, M.~E. Cates,
  J.~Vollmer, A.~Wagner and D.~Vollmer, \emph{Philosophical Transactions of the
  Royal Society of London. Series A: Mathematical, Physical and Engineering
  Sciences}, 2003, \textbf{361}, 793--807\relax
\mciteBstWouldAddEndPuncttrue
\mciteSetBstMidEndSepPunct{\mcitedefaultmidpunct}
{\mcitedefaultendpunct}{\mcitedefaultseppunct}\relax
\EndOfBibitem
\bibitem[Singh \emph{et~al.}(2011)Singh, Mukherjee, Vermeulen, Barkema, and
  Puri]{PuriControl}
A.~Singh, A.~Mukherjee, H.~M. Vermeulen, G.~T. Barkema and S.~Puri, \emph{The
  Journal of Chemical Physics}, 2011, \textbf{134}, 044910\relax
\mciteBstWouldAddEndPuncttrue
\mciteSetBstMidEndSepPunct{\mcitedefaultmidpunct}
{\mcitedefaultendpunct}{\mcitedefaultseppunct}\relax
\EndOfBibitem
\bibitem[Yan and Xie(2008)]{Lamellar}
L.-T. Yan and X.-M. Xie, \emph{The Journal of Chemical Physics}, 2008,
  \textbf{128}, 034901\relax
\mciteBstWouldAddEndPuncttrue
\mciteSetBstMidEndSepPunct{\mcitedefaultmidpunct}
{\mcitedefaultendpunct}{\mcitedefaultseppunct}\relax
\EndOfBibitem
\bibitem[Fialkowski and Holyst(2002)]{Holyst2002}
M.~Fialkowski and R.~Holyst, \emph{The Journal of Chemical Physics}, 2002,
  \textbf{117}, 1886--1892\relax
\mciteBstWouldAddEndPuncttrue
\mciteSetBstMidEndSepPunct{\mcitedefaultmidpunct}
{\mcitedefaultendpunct}{\mcitedefaultseppunct}\relax
\EndOfBibitem
\bibitem[Podariu and Chakrabarti(2007)]{Podariu2007}
I.~Podariu and A.~Chakrabarti, \emph{The Journal of Chemical Physics}, 2007,
  \textbf{126}, 154509\relax
\mciteBstWouldAddEndPuncttrue
\mciteSetBstMidEndSepPunct{\mcitedefaultmidpunct}
{\mcitedefaultendpunct}{\mcitedefaultseppunct}\relax
\EndOfBibitem
\bibitem[Henderson and Clarke(2004)]{ClarkeTwoStep}
I.~C. Henderson and N.~Clarke, \emph{Macromolecules}, 2004, \textbf{37},
  1952--1959\relax
\mciteBstWouldAddEndPuncttrue
\mciteSetBstMidEndSepPunct{\mcitedefaultmidpunct}
{\mcitedefaultendpunct}{\mcitedefaultseppunct}\relax
\EndOfBibitem
\bibitem[Li \emph{et~al.}(2011)Li, Shi, Wang, Liu, and
  Wang]{WangLi2011ViscousTwoStep}
Y.~C. Li, R.~P. Shi, C.~P. Wang, X.~J. Liu and Y.~Wang, \emph{Physical Review
  E}, 2011, \textbf{83}, 041502\relax
\mciteBstWouldAddEndPuncttrue
\mciteSetBstMidEndSepPunct{\mcitedefaultmidpunct}
{\mcitedefaultendpunct}{\mcitedefaultseppunct}\relax
\EndOfBibitem
\bibitem[Hashimoto \emph{et~al.}(2000)Hashimoto, Hayashi, and
  Jinnai]{HashimotoExperimental1}
T.~Hashimoto, M.~Hayashi and H.~Jinnai, \emph{The Journal of Chemical Physics},
  2000, \textbf{112}, 6886--6896\relax
\mciteBstWouldAddEndPuncttrue
\mciteSetBstMidEndSepPunct{\mcitedefaultmidpunct}
{\mcitedefaultendpunct}{\mcitedefaultseppunct}\relax
\EndOfBibitem
\bibitem[Hayashi \emph{et~al.}(2000)Hayashi, Jinnai, and
  Hashimoto]{HashimotoExperimental2}
M.~Hayashi, H.~Jinnai and T.~Hashimoto, \emph{The Journal of Chemical Physics},
  2000, \textbf{112}, 6897--6909\relax
\mciteBstWouldAddEndPuncttrue
\mciteSetBstMidEndSepPunct{\mcitedefaultmidpunct}
{\mcitedefaultendpunct}{\mcitedefaultseppunct}\relax
\EndOfBibitem
\bibitem[Tanaka(1994)]{TanakaPRLExperimental}
H.~Tanaka, \emph{Physical Review Letters}, 1994, \textbf{72}, 3690--3693\relax
\mciteBstWouldAddEndPuncttrue
\mciteSetBstMidEndSepPunct{\mcitedefaultmidpunct}
{\mcitedefaultendpunct}{\mcitedefaultseppunct}\relax
\EndOfBibitem
\bibitem[Sigehuzi and Tanaka(2004)]{Tanaka2004Experimental}
T.~Sigehuzi and H.~Tanaka, \emph{Physical Review E}, 2004, \textbf{70},
  051504\relax
\mciteBstWouldAddEndPuncttrue
\mciteSetBstMidEndSepPunct{\mcitedefaultmidpunct}
{\mcitedefaultendpunct}{\mcitedefaultseppunct}\relax
\EndOfBibitem
\bibitem[Plapp and Gouyet(1997)]{PhysRevLett.78.4970}
M.~Plapp and J.-F. Gouyet, \emph{Physical Review Letters}, 1997, \textbf{78},
  4970--4973\relax
\mciteBstWouldAddEndPuncttrue
\mciteSetBstMidEndSepPunct{\mcitedefaultmidpunct}
{\mcitedefaultendpunct}{\mcitedefaultseppunct}\relax
\EndOfBibitem
\bibitem[Plapp and Gouyet(1999)]{Plapp99}
M.~Plapp and J.-F. Gouyet, \emph{The European Physical Journal B - Condensed
  Matter and Complex Systems}, 1999, \textbf{9}, 267--282\relax
\mciteBstWouldAddEndPuncttrue
\mciteSetBstMidEndSepPunct{\mcitedefaultmidpunct}
{\mcitedefaultendpunct}{\mcitedefaultseppunct}\relax
\EndOfBibitem
\bibitem[Pagonabarraga and Cates(2003)]{ignacio}
I.~Pagonabarraga and M.~E. Cates, \emph{Macromolecules}, 2003, \textbf{36},
  934--949\relax
\mciteBstWouldAddEndPuncttrue
\mciteSetBstMidEndSepPunct{\mcitedefaultmidpunct}
{\mcitedefaultendpunct}{\mcitedefaultseppunct}\relax
\EndOfBibitem
\bibitem[Note1()]{Note1}
More specifically, to determine the colour of a site $i$, we blended together
  the colours red, green, and blue. The intensity of each of these colours at a
  given site varies from $0$ to $1$. In our scheme, red, green, and blue
  intensities are given by $1-p_i^A,p_i^0,1-p_i^B$, respectively. (Remember the
  notation for the local concentration of vacancies, i.e.\
  $p_i^0=1-p_i^A-p_i^B$.) This leads to the colour key shown in the top-left
  part of Fig.\ 8 in ref.\ \citenum {PabloPeter1}. It is plotted in triangular
  colour space in $(p^A, p^B, 1 - p^A - p^B)$, dropping the site index $i$. For
  example, if the concentration of particles of species $A$ at one site is high
  (low), and the concentration of particles of species $B$ at the same site is
  low (high), then the site colour will tend towards blue (red); if the
  concentrations of all species are all low, then the site colour will tend to
  white.\relax
\mciteBstWouldAddEndPunctfalse
\mciteSetBstMidEndSepPunct{\mcitedefaultmidpunct}
{}{\mcitedefaultseppunct}\relax
\EndOfBibitem
\bibitem[Li \emph{et~al.}(2017)Li, Choi, and Kim]{li2017computationally}
Y.~Li, Y.~Choi and J.~Kim, \emph{Computers and Mathematics with Applications},
  2017, \textbf{73}, 1855--1864\relax
\mciteBstWouldAddEndPuncttrue
\mciteSetBstMidEndSepPunct{\mcitedefaultmidpunct}
{\mcitedefaultendpunct}{\mcitedefaultseppunct}\relax
\EndOfBibitem
\bibitem[Note2()]{Note2}
This has been performed using a function from \textit {Mathematica 11.2} called
  `ComponentMeasurements', which has been applied to a binary image obtained
  from our data; to construct the binary image we defined that total densities
  below $0.5$ constitute gas bubbles. Therefore the interfaces are mostly
  considered part of the liquid.\relax
\mciteBstWouldAddEndPunctfalse
\mciteSetBstMidEndSepPunct{\mcitedefaultmidpunct}
{}{\mcitedefaultseppunct}\relax
\EndOfBibitem
\bibitem[Note3()]{Note3}
For the $w_s=1$ data, in order for this to be true we needed to use a slightly
  smaller region rather than the entire system.\relax
\mciteBstWouldAddEndPunctfalse
\mciteSetBstMidEndSepPunct{\mcitedefaultmidpunct}
{}{\mcitedefaultseppunct}\relax
\EndOfBibitem
\bibitem[Lifshitz and Slyozov(1961)]{LIFSHITZ196135}
I.~Lifshitz and V.~Slyozov, \emph{Journal of Physics and Chemistry of Solids},
  1961, \textbf{19}, 35 -- 50\relax
\mciteBstWouldAddEndPuncttrue
\mciteSetBstMidEndSepPunct{\mcitedefaultmidpunct}
{\mcitedefaultendpunct}{\mcitedefaultseppunct}\relax
\EndOfBibitem
\bibitem[Huse(1986)]{Huse1986}
D.~A. Huse, \emph{Physical Review B}, 1986, \textbf{34}, 7845--7850\relax
\mciteBstWouldAddEndPuncttrue
\mciteSetBstMidEndSepPunct{\mcitedefaultmidpunct}
{\mcitedefaultendpunct}{\mcitedefaultseppunct}\relax
\EndOfBibitem
\bibitem[Streitenberger(2013)]{STREITENBERGER20135026}
P.~Streitenberger, \emph{Acta Materialia}, 2013, \textbf{61}, 5026 --
  5035\relax
\mciteBstWouldAddEndPuncttrue
\mciteSetBstMidEndSepPunct{\mcitedefaultmidpunct}
{\mcitedefaultendpunct}{\mcitedefaultseppunct}\relax
\EndOfBibitem
\bibitem[Wagner(1961)]{wagner1961theorie}
C.~Wagner, \emph{Zeitschrift f{\"u}r Elektrochemie, Berichte der
  Bunsengesellschaft f{\"u}r physikalische Chemie}, 1961, \textbf{65},
  581--591\relax
\mciteBstWouldAddEndPuncttrue
\mciteSetBstMidEndSepPunct{\mcitedefaultmidpunct}
{\mcitedefaultendpunct}{\mcitedefaultseppunct}\relax
\EndOfBibitem
\bibitem[Yan and Xie(2006)]{TangMingSurfaceEffects}
L.-T. Yan and X.-M. Xie, \emph{Macromolecules}, 2006, \textbf{39},
  2388--2397\relax
\mciteBstWouldAddEndPuncttrue
\mciteSetBstMidEndSepPunct{\mcitedefaultmidpunct}
{\mcitedefaultendpunct}{\mcitedefaultseppunct}\relax
\EndOfBibitem
\bibitem[Tafa \emph{et~al.}(2001)Tafa, Puri, and Kumar]{tafa2001kinetics}
K.~Tafa, S.~Puri and D.~Kumar, \emph{Physical Review E}, 2001, \textbf{64},
  056139\relax
\mciteBstWouldAddEndPuncttrue
\mciteSetBstMidEndSepPunct{\mcitedefaultmidpunct}
{\mcitedefaultendpunct}{\mcitedefaultseppunct}\relax
\EndOfBibitem
\end{mcitethebibliography}
\bibliographystyle{rsc} 

\end{document}